\documentstyle[preprint,aps,epsf]{revtex}
\tighten
\begin{document}
\draft

\title{Theoretical Aspects of Science with Radioactive Nuclear
Beams\footnote{Will be published in {\em Theme Issue on Science
with Beams of Radioactive Nuclei}, Philosophical Transactions, 
ed. by W. Gelletly}}

\author{
Jacek Dobaczewski$^{1-3}$ and
 Witold Nazarewicz$^{1,3,4}$}

\address {
$^1$Department of Physics, University of Tennessee,
 Knoxville, TN 37996, U.S.A.\\
$^2$Joint Institute for Heavy Ion Research,
   Oak Ridge National Laboratory                                \\[-1mm]
   P.O. Box 2008, Oak Ridge,   TN 37831, U.S.A.                 \\
$^3$Institute of Theoretical Physics, Warsaw University         \\[-1mm]
   Ho\.za 69,  PL-00681, Warsaw, Poland                         \\
$^4$Physics Division, Oak Ridge National Laboratory             \\[-1mm]
   P.O. Box 2008, Oak Ridge,   TN 37831, U.S.A.
}
\maketitle

\begin{abstract}

Physics of radioactive nuclear beams is one of the main
frontiers of nuclear science  today.  Experimentally, thanks to
technological developments, we are on the verge of invading the
territory of extreme $N/Z$ ratios  in an unprecedented way.
Theoretically,  {nuclear exotica} represent a formidable
challenge for the nuclear many-body theories and their power to predict
nuclear properties in nuclear terra incognita.
It is important to remember
 that
the lesson learned by going to the limits of the nuclear
binding is also important for ``normal" nuclei from  the
neighborhood of the beta stability valley. And, of course,
radioactive nuclei are crucial astrophysically; they pave the
highway along which the nuclear material is  transported
up in the proton and neutron numbers
 during the complicated synthesis process in
stars.

\end{abstract}

\narrowtext

\section{Introduction}

\newif\iffirstfig \global\firstfigfalse

There are only   263 stable nuclei; they
are surrounded by  radioactive ones.
Some
of the unstable nuclei
 are long-lived and can be found on Earth,  some are
man-made (actually, as many as  $\sim$2,200 nuclei have
been produced in nuclear laboratories), and several
thousand
nuclei are the yet-unexplored exotic species.
 The decay characteristics
of most radioactive  nuclei are
determined by the beta decay, i.e., by weak interactions.
For heavier nuclei, where
the  electromagnetic interaction plays a more important
role, other decay channels, such as
emission of alpha particles  or
spontaneous fission,  dominate.
Moving away from
stable nuclei by adding either protons or neutrons,
one finally reaches the
particle drip
lines.  The nuclei beyond the
drip lines are unbound to nucleon emission; that is,
for those systems the
strong interaction is unable to bind $A$ nucleons as
one nucleus.

So, the territory of exotic nuclei is enormous.
The  uncharted regions of  the ($N$, $Z$) plane
contain information that can
answer  many questions of fundamental
importance for science:
How many
protons and neutrons can be clustered  together by  the strong
interaction to form a bound nucleus?
What are the proton and neutron magic
numbers of the  exotic nuclei?
What are the properties of very short-lived
exotic nuclei with extreme neutron-to-proton ratio
$N/Z$? What is the effective nucleon-nucleon interaction
in the nucleus having a very large neutron excess?
There
are also related questions in the field of nuclear astrophysics.
Since radioactive nuclei are produced in many astrophysical
sites, knowledge of their properties is crucial to the
understanding of the underlying processes.

Nuclear life  far from stability is different from that
around the stability line;
the promised
access to  completely new
combinations of proton and neutron numbers offers  prospects
for new structural phenomena.
The unique  structural factor is the weak binding; hence
closeness to the particle continuum.
The main
objective of this paper is to
discuss some of the theoretical challenges and
opportunities of research with exotic nuclear beams.

The paper is organized as follows. Section~\ref{theory} contains
a  brief review of theoretical developments related to the
physics of exotic nuclei.
Sections~\ref{largeNZ} and \ref{smallNZ}
contain some of
the physics issues of the neutron and proton drip lines, respectively,
including a discussion on the theoretical uncertainties in
determining the  particle drip lines.
Finally, conclusions are contained in Sec.~\ref{conclusions}.

\section{New Theoretical Aspects of Physics with Exotic Beams}\label{theory}

From a theoretical point of view,
spectroscopy of exotic nuclei
offers a unique test of those components of
effective interactions that
depend on the isospin degrees of freedom.
In principle, the effective nucleon-nucleon interaction
in   heavy nuclei should be
obtained by means of the Br\"uckner renormalization
which corrects the free interaction for
the effects due to the
nuclear  medium. In practice, however, the effective  interaction
is approximated by means of some
phenomenological density-dependent force
with parameters that are  usually fitted to stable nuclei
and to selected  properties of the infinite nuclear matter.
Hence,  it is by no means obvious that
the isotopic trends far from stability
predicted by commonly used effective interactions
are correct. In the models
aiming at such an extrapolation,
the important  questions asked are:
What is the density dependence of the two-body central force
\cite{[Dab77],[Fre86],[Pea94]}?
What is the density and radial dependence of the one-body spin-orbit force
\cite{[Sha95],[Rei94],[Cha95],[Ons97]}?
Does the spin-orbit splitting strongly vary with $N/Z$
\cite{[Pud96]}?
What is the form of the pairing interaction in weakly bound nuclei
\cite{[Ber91],[Dob96],[Fay96],[She96]}?
What is the importance of the effective mass (i.e.,
the non-locality of the force)
for isotopic trends  \cite{[Dob95c]}?
What is the role of the medium effects
(renormalization) and of the core polarization in the nuclear exterior
(halo or skin region) where the nucleonic density is small
\cite{[Kuo97]}?
Similar questions are asked in connection  with  properties
of nuclear matter \cite{[Cha95],[Cug87],[Bro88a],[Gmu92]},
neutron droplets \cite{[Pud96]}, and the
 physics of the neutron-star crust
\cite{[Pet95],[Pet95a]}.

The  radioactive nuclear beams experimentation is expected to expand the
range of nuclei known. That is,
by going to  nuclei with extreme $N/Z$ ratios, one can
magnify the isospin-dependent terms of the effective
 interaction
(which are small in ``normal" nuclei). The hope is that after
probing  these terms at the
limits of extreme isospin, we can later go back  to the valley of
stability and   improve the description of ``normal" nuclei.

But this task is not going to be easy. In many respects, weakly bound nuclei
are much more difficult to treat theoretically than well-bound systems.
Hence, before tackling  the problem of force parametrization
at the extremes, one should be sure that the applied
theoretical tools of the nuclear many-body problem are appropriate.

As mentioned above, the main theoretical
challenge is the correct treatment of the particle continuum.
For
weakly bound nuclei, the Fermi energy lies very close to zero, and
the decay channels
must be taken into account explicitly.
As a result,
many cherished approaches of nuclear theory such as the
conventional  shell model, the
pairing theory, or the macroscopic-microscopic approach
must be modified. But there  is also  a splendid opportunity:
the explicit coupling between bound states and continuum,
and the presence of low-lying scattering states
invite  strong interplay and cross-fertilization
between
nuclear structure and reaction theory. Many methods developed
by  reaction theory can now be applied to structure aspects
of loosely bound systems.

How to extend  traditional tools of nuclear theory
to account for  the scattering of nucleons from bound single-particle
orbitals to unbound states?
The closeness of particle continuum reverberates in two aspects
of the theoretical description. Firstly, the particles forming a bound
nuclear state can effortlessly {\em virtually} scatter back and forth
into the particle continuum phase space. This process must conserve
the compactness of the nuclear wave function which remains bound
even with such a virtual scattering taken into account. A theoretical
description of this kind of effects still remains a virgin territory,
although some progress has been made in the analysis of the virtual
pair scattering \cite{[Dob84],[Dob96]}.
Secondly, nucleons can very easily leave the nucleus
altogether and enter the particle continuum through the {\em real} scattering.
For this, it is enough to slightly shake the nucleus by providing it
with a little bit of energy.
This is an old problem which, in the context
of excited states near or above the particle threshold,
has been addressed by
the continuum shell model (CSM)
\cite{[Fan61],[Glo67],[Iba70],[Phi77a],[Bar77],[Mic78],[Hal80],[Isk91]}.
In the CSM, the continuum
states (decay channels)  and bound states are treated  on  equal footing.
 Consequently,  correlations due to the coupling to
resonances, the spatial extension effects in weakly bound states,
the structure of resonances, and
the structure of particle transfer form factors
are properly described by the CSM. So far, most applications of the CSM
have been
concerned with the situation when there is only one particle
occupying the
shell-model continuum. This is because  the
continuum-continuum coupling is difficult to treat \cite{[Wen87]}.

Often, particle continuum is approximated by
the {\em quasibound states}, i.e., the states
resulting from the diagonalization of a finite potential in a large
basis \cite{[Bol72],[Naz94]} or by enclosing
the finite nuclear potential within an infinite
well with walls positioned at a large distance from the nuclear surface
\cite{[Ben96],[Ghi96]}.
More sophisticated methods of discretizing continuum include
the Sturmian function expansions and resonant state expansions.
Sturmian functions, also known as Weinberg states,
form a discrete set of states which behave asymptotically
as outgoing waves. They have been used as a basis in the
solution of scattering equations, including various applications of
the CSM
\cite{[Glo67],[Vaa79],[Raw82],[Bub91],[Rid97]}.
The Gamow states are eigenstates
of the time-independent Schr\"odinger equation
 with complex eigenvalues \cite{[Rom72],[Ber68],[Ver87]}. They have been
applied to many problems involving an unbound spectrum
\cite{[Lin94],[Ber96],[For97]}.

In the  description of weakly bound systems,
pairing interaction plays a unique role (see Section \ref{pairing} below).
In the BCS or BCS-like methods
based on bound and  quasibound states (for examples
of such calculations see, e.g., Refs. \cite{[Naz94],[San97]}),
the virtual scattering of nucleonic pairs
from bound states to the positive-energy states
leads to the presence of
a ``particle gas" surrounding the nucleus \cite{[Dob84]}.
To show it, we decompose the BCS wave function into
contributions from
bound states ($\epsilon_i<0$) and  quasibound states ($\epsilon_j>0$):
\begin{equation}\label{BCSW}
\Psi_{BCS} = \prod_{i, \epsilon_i<0}
\left( U_i + V_i a^\dagger_i a^\dagger_{\bar i}\right)
\prod_{j, \epsilon_j>0}
\left(U_j + V_j a^\dagger_j a^\dagger_{\bar j}\right)
|0\rangle.
\end{equation}
While the bound-state component  in Eq.~(\ref{BCSW}) represents
the localized wave function, i.e., it decays
asymptotically, the second part represents the contribution from
 quasibound states  and  leads
to non-localized  densities
with incorrect asymptotic behavior. Indeed,
although the nuclear densities eventually vanish
at large distances by construction (finite size of the basis, finite
size of the box in which calculations are performed),
the  wave functions  of positive-energy states
do not decay outside
 the nuclear volume. As discussed below in Sec.~\ref{pairing},
this problem is overcome in the
Hartree-Fock-Bogolyubov (HFB) method with a realistic
pairing interaction in which the coupling of bound states to
the particle continuum is correctly taken into account \cite{[Dob84],[Dob96]}.

\begin{figure}[tbh]
\epsfxsize 15.cm
\centerline{\epsfbox{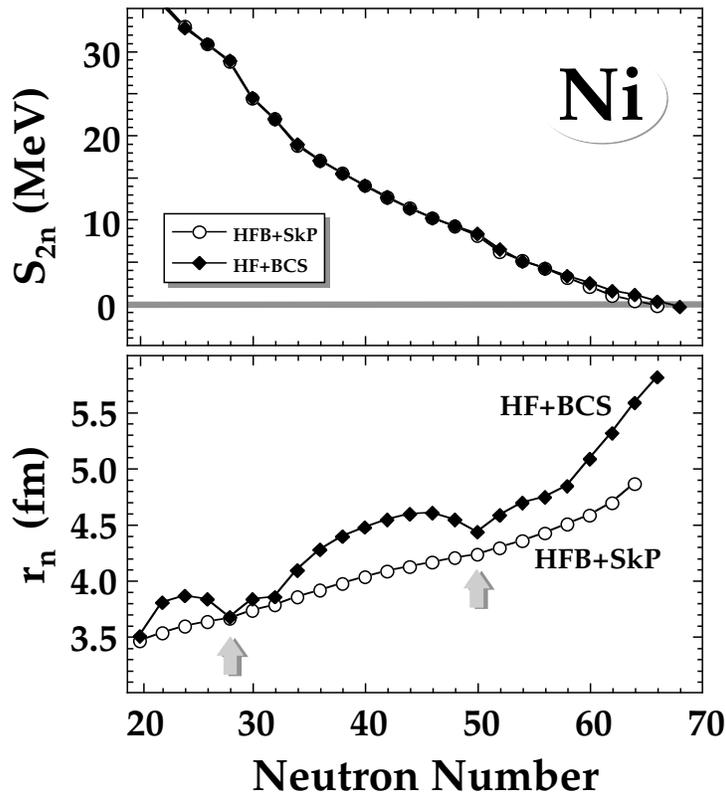}}
\caption{
Top: Two-neutron
 separation energies for the
even-even
nickel isotopes  predicted in the
HFB and HF+BCS calculations with the
 Skyrme
interaction SkP.
Bottom:  The predicted rms neutron radii.
The arrows indicate the neutron subshell closures at $N$=28 and  50.
(From Ref.\ \protect\cite{[Dob96a]}.)
}
\label{Nickel}
\end{figure}
To illustrate this point, critical in the context of
calculations for drip-line nuclei, Fig.~\ref{Nickel} displays
theoretical two-neutron
 separation energies, $S_{2n}$, and the rms neutron radii for the
even-even
nickel isotopes obtained in the
HFB and HF+BCS calculations.
In the HF+BCS variant, the self-consistent
pairing gaps obtained from the HFB calculations were used
within the fixed-gap approximation.
As seen in Fig.~\ref{Nickel},
the values of $S_{2n}$ obtained in the HFB and HF+BCS calculations
agree very well; some deviations are
seen only for the neutron drip-line
systems with $N$$>$60, where the HF+BCS values are slightly lower.
However, this excellent agreement does not extend to
neutron radii. In the  HFB calculations
the neutron
radii behave very smoothly as a function of $N$.
On the other hand,
in the HF+BCS model
there is a dramatic increase in the neutron radii
between magic numbers and
for the neutron-rich nuclei
resulting from the unphysical occupation of positive-energy
quasibound states.
The difference between values of radii obtained within the
HFB and HF+BCS calculations
can be as large as 0.8\,fm and increases dramatically
for weakly bound systems.
Other examples illustrating the unphysical effect of
the particle gas  can be found in
Refs.~\cite{[Dob96],[Naz96]}.

Consequently, for large exotic nuclei, the self-consistent HFB treatment
is not a matter of choice, it is a must. The calculations
are not easy, especially if the self-consistent symmetries (e.g.,
spherical symmetry) are broken.
Possible strategies for solving the HFB equations  include
the two-step diagonalization \cite{[Ter96]}, the
gradient method in the canonical representation \cite{[Muh84]},
or the state-dependent Hamiltonian method \cite{[Rei97]}.
Some  examples
of HFB calculations are discussed in the following sections. Other
calculations with
self-consistent inclusion of pairing and continuum states
include:
quasiclassical Lagrangian method calculations in the coordinate
representation
\cite{[Kho82],[Zve84],[Zve85],[Smi88],[Zve91]} and
application of the
relativistic Hartree-Bogolyubov theory
in coordinate space
to light nuclei \cite{[Men96],[Poe97]}.

In order to describe excited states, one has to go beyond the mean-field
approximation. One method, often used in the description of low-lying collective
states in drip-line nuclei, is the
continuum random phase approximation (CRPA)
 based on
the single-particle Green function approach in the coordinate
representation \cite{[Mig67],[Shl75]}. In particular, the
  CRPA
has been employed extensively to the very low-energy
multipole  strength  in  drip-line nuclei
\cite{[Kam93],[Sag96],[Ham96],[Ham96a]}. However,
for meaningful
predictions of excited states in weakly bound nuclei where pairing
is expected to be very important, it is necessary to use
 the quasiparticle RPA scheme based
on the  coordinate space HFB formalism. Calculations
along these lines can be found in recent Refs.~\cite{[Bor95],[Bor96]}.

The following sections contain a brief
description of selected
highlights of the physics of radioactive nuclear beams.
It is to be noted that there are many other important topics that
have been left out in our discussion  (e.g., physics of very light
drip-line nuclei,
physics of beta decay, reaction aspects involving radioactive ions). A key
point is that the variety of exciting new
phenomena is one of the driving forces behind
research with exotic beams.

\section{Physics of Neutron-Rich Nuclei: upper limits of the $N/Z$ Ratio}
\label{largeNZ}

Since neutrons  do not
carry an electric charge and  do not repel each
other, many neutrons can be added to nuclei starting from the
valley of stability. As a result,
the  ``lever arm'' separating the neutron drip line from the valley of
stability is large and difficult to probe experimentally;
 except for the lightest nuclei,
the bounds of the neutron stability are not known.
But it is just for the  nuclei with the extreme neutron excess
that theory predicts many new and initially unexpected phenomena.
In addition to nuclear structure interest,
the neutron-rich environment   is  important
for astrophysics and cosmology.

\subsection{Density distributions}
\label{densities}

Neutron-rich nuclei are characterized by spatially extended density distributions
which give rise to large nuclear radial moments. Extreme cases are
halo nuclei -- loosely bound few-body systems with about thrice more
neutrons than protons.
The halo region is a zone of weak
binding in which quantum effects play a critical role in distributing
nuclear density in regions not classically allowed.

 In the heavier,
neutron-rich  nuclei, where  the concept of mean field
 is better applicable,
the separation into a ``core" and ``valence
nucleons" seems less justified.
However, also in these nuclei
the weak neutron binding implies
the existence of the neutron skin
(i.e., a dramatic excess of neutrons at large distances).

\begin{figure}[tbh]
\epsfxsize 15.cm
\centerline{\epsfbox{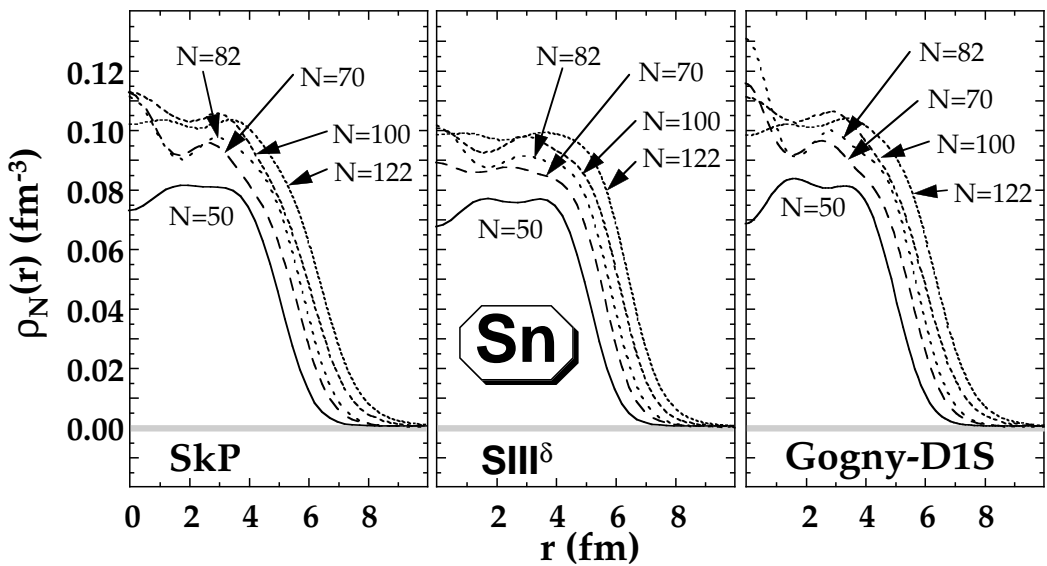}}
\vspace*{-3cm}
\caption{
Self-consistent spherical neutron densities $\rho_N(r)$
calculated with the SkP,
SIII$^\delta$, and D1S interactions for
selected tin isotopes across the $\beta$-stability valley.
Since proton and neutron densities in the nucleus $^{100}$Sn
are very similar \protect\cite{[Dob94]}, the deviation from the neutron
density at $N$=50 roughly represents the skin effect.
(From Ref.~\protect\cite{[Dob96]}.)
}
\label{dens}
\end{figure}
Figure~\ref{dens} displays the  neutron HFB densities for
several tin isotopes across the stability valley, calculated
with the effective interactions SkP\cite{[Dob84]},
SIII$^\delta$ \cite{[Dob95c]}, and D1S \cite{[Dec80]}.
The densities obtained with these forces
are qualitatively very similar. One can see that
adding  neutrons results in a simultaneous increase of the
central neutron density, and of the density in the surface
region.
The relative magnitude of the two effects is governed by
a balance between the volume and the surface asymmetry energies
of effective interactions. Since all three forces  considered
have been fitted in a similar way to bulk nuclear properties
(including the isospin dependence), the resulting balance between
the volume and the surface isospin effects is similar. Of
course, this does not exclude some differences which are seen
when a more detailed comparison is carried out.
As will be seen  in Sec~\ref{pairing} below,
 pair densities depend much stronger on the effective forces.

\begin{figure}[tbh]
\epsfxsize 15.cm
\centerline{\epsfbox{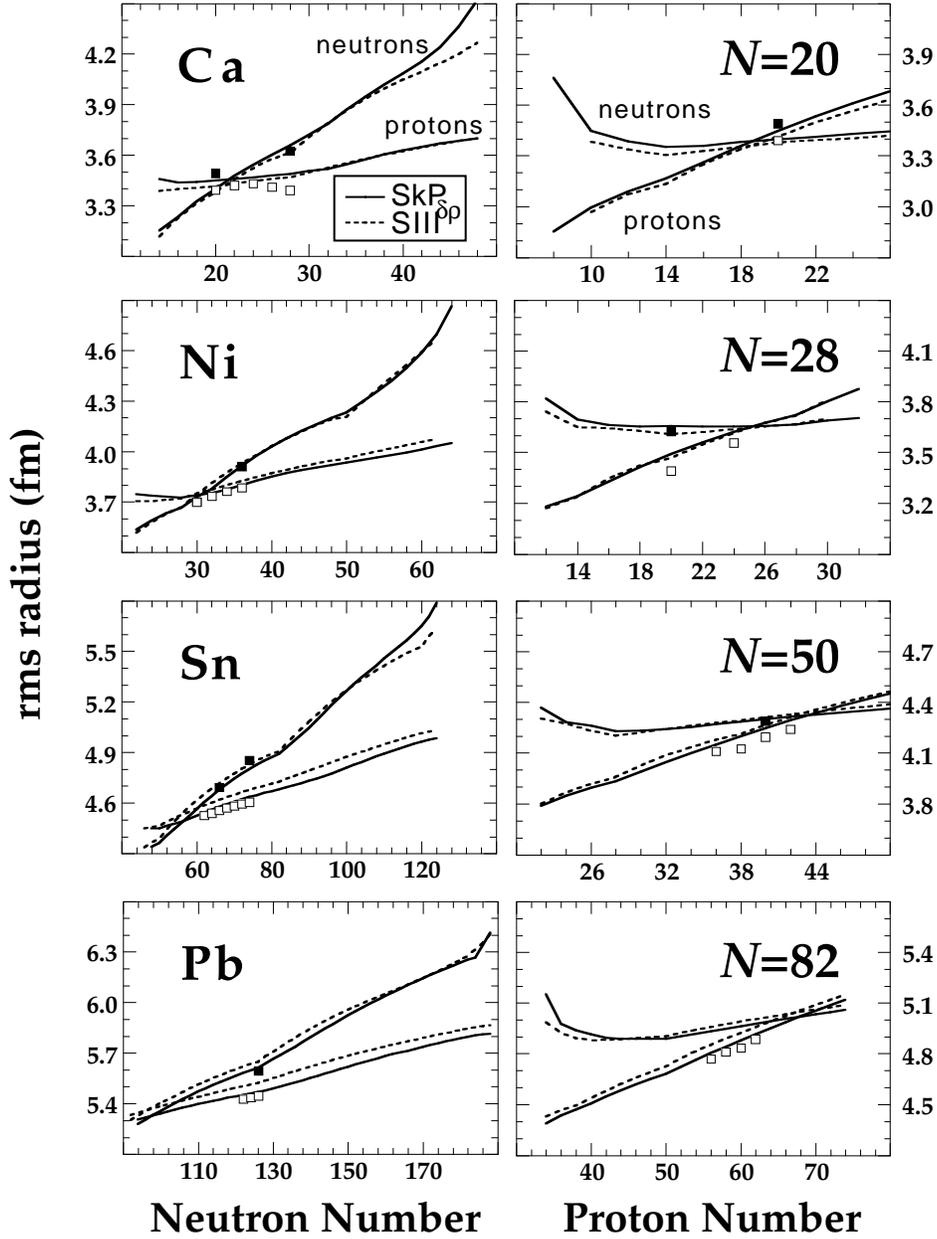}}
\vspace*{-2.5cm}
\caption{
Comparison of experimental (symbols) and calculated
(HFB+SkP, solid line; HFB+SIII$^{\delta\rho}$, dashed line)
 proton and
neutron rms radii for the semi-magic nuclei
with $N$\,(or $Z$)=20, 28, 50, and 82. The experimental proton radii
were extracted from experimental charge radii \protect\cite{[Nad94]}.
The experimental neutron
radii were obtained from the analysis of the high-energy
proton-nucleus scattering \protect\cite{[Bat89]}.
(From Ref.\ \protect\cite{[Dob96a]}.)
}
\label{Rexpt}
\end{figure}
A very interesting aspect of nuclei far from stability is an
increase in their radial dimension with decreasing particle
separation energy \cite{[Rii92]}.  This effect is especially strong
in weakly bound nuclei close to the neutron drip line.
Figure~\ref{Rexpt} presents the
comparison of experimental (symbols) and calculated
proton and
neutron rms radii for the semi-magic isotopes and isotones
with $Z$\,(or $N$)=20, 28, 50, and 82.
For neutron radii in general,
and for proton radii in light and medium-mass nuclei, the
 HFB+SkP and HFB+SIII$^{\delta\rho}$
models yield very similar predictions.
In heavy nuclei ($Z$ or $N$=50, 82)
the proton radii in SkP are lower than in
SIII$^{\delta\rho}$, in closer agreement with
 experiment.
In Fig.~\ref{Rexpt}, the neutron skin
is manifested by a rapid increase in
neutron rms radii  when approaching the two-neutron drip line.
This effect is very localized in neutron number; it appears
only in a few  nuclei
in the immediate neighborhood of the neutron drip line.

\subsection{Pairing correlations}\label{pairing}

Pairing correlations play a very special role in drip-line nuclei
\cite{[Dob84],[Dob96]}.
This is seen from the approximate HFB
relation  between the Fermi level
$\lambda$, pairing gap $\Delta$, and the particle
separation energy $S\approx -\lambda - \Delta$.
At the drip line  $S$ is very small and
$\lambda + \Delta \sim 0$. Consequently,
the single-particle
field characterized by $\lambda$
and the pairing field $\Delta$
are equally important. In other words, contrary to the situation
encountered close to the line of beta stability, the pairing component
of the Hamiltonian
can no longer be treated as a {residual} interaction, i.e., a
small perturbation
important only in the neighborhood of the Fermi surface.

Surprisingly, very little is known about the pairing
channel of the
effective nucleon-nucleon interaction.
In most calculations, the pairing Hamiltonian
has been approximated by the  state-independent seniority
pairing force, or schematic multipole pairing interaction
\cite{[Lan64]}. Such oversimplified forces, usually treated by
means of the BCS approximation, perform remarkably well when
applied to nuclei in the neighborhood of the stability valley
(where, as pointed out above,  pairing can be considered as a
small correction). As a result, considerable effort was devoted
in the past to optimizing the HF part of the interaction, while
leaving the pairing  component aside.

A detailed discussion of the present status of
effective interactions in the particle-particle
channel can be found in Ref.~\cite{[Dob96]}.
The main questions pertaining to this problem are:
What is the microscopic origin of the pairing interaction
\cite{[Bru60],[Eme60],[Del95],[Kuc91],[Kad87]}?
What is  the
role of finite range and the importance of density
dependence \cite{[Sap65],[Zaw87],[Reg88],[Fay94],[Bul80],[Bel87]}?
How can properties of the  pairing force  be tested experimentally?
These questions are
 of  considerable importance
not only for nuclear physics but also for nuclear astrophysics
and cosmology \cite{[Pet95],[Von91],[Deb94]}.

\begin{figure}[tbh]
\epsfxsize 15.cm
\centerline{\epsfbox{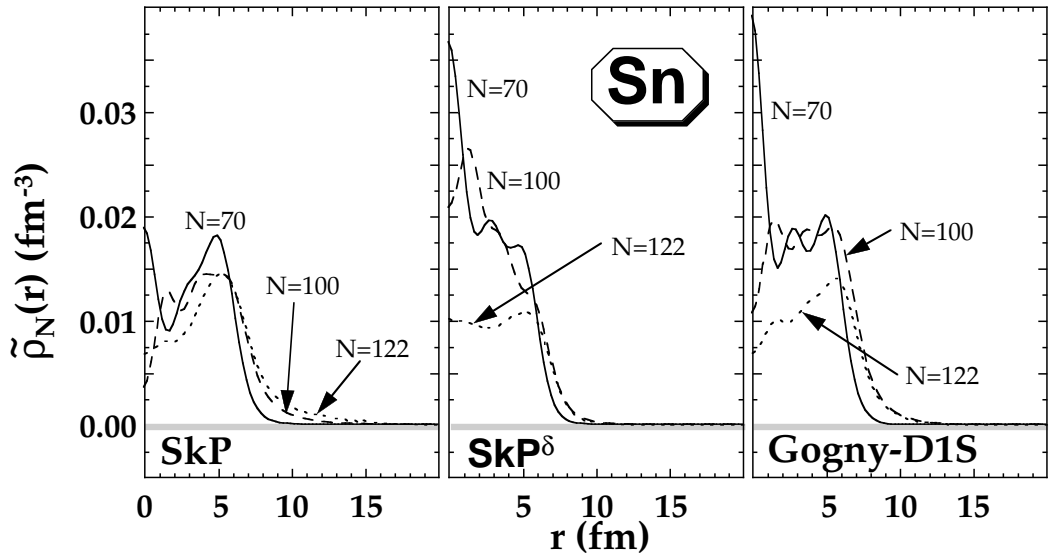}}
\vspace*{-3cm}
\caption{
Self-consistent spherical
neutron pairing densities $\tilde\rho_N(r)$ calculated
with the
SkP, SkP$^\delta$, and D1S interactions for
selected tin isotopes across the $\beta$-stability valley.
(From Ref.~\protect\cite{[Dob96]}.)
}
\label{densitiesHFB}
\end{figure}

Because of strong surface effects, the properties of weakly bound nuclei
are perfect laboratories in which to study
the density dependence of pairing interactions.
As an example of what can be expected far from stability,
Fig.~\ref{densitiesHFB} displays  the neutron HFB
local pairing  densities
$\tilde\rho(r)$
calculated for several tin isotopes across the stability
valley, and for three different effective interactions:
Skyrme interactions SkP and SkP$^\delta$, and the finite-range Gogny
interaction D1S.
[The density $\tilde\rho(\bbox{r})$ is proportional to the probability of finding
the correlated pair of nucleons at point $\bbox{r}$,
see Ref.~\cite{[Dob96]} for definitions and discussion.]
The pairing densities shown in Fig.\ \ref{densitiesHFB}
nicely reflect different
characters of the interactions used. Namely,
the contact force SkP$^\delta$ leads to pairing
densities that are, in general, largest at the origin and decrease
towards the surface;
this is  characteristic of the volume-type pairing correlations.
A different pattern appears for the SkP results, where the density
dependence renders the pairing  interaction strongly peaked
at the surface. In this case, the pairing densities tend to increase
when going from the center
of the nucleus towards its surface.
A more pronounced dependence on the neutron excess is
seen here in the surface region. Near the drip line, the pairing
density develops a long tail extending towards large distances.
The results obtained for the finite-range Gogny interaction
exhibit features intermediate between surface and volume-type
pairing correlations.

An experimental observable that may probe the character of the
pairing field is the pair transfer form factor, directly related
to the pairing  density $\tilde\rho$. The difference in the
asymptotic behavior of single-particle density $\rho$ and  pair
density $\tilde\rho$ in a weakly bound system
can be probed by comparing
the energy dependence of one-particle and pair-transfer cross
sections.  Such measurements, when performed for both stable and
neutron-rich nuclei, can shed some light on the asymptotic
properties of the pair densities; hence on the character of the pairing
field.

\begin{figure}[htb]
\epsfxsize 10.cm
\centerline{\epsfbox{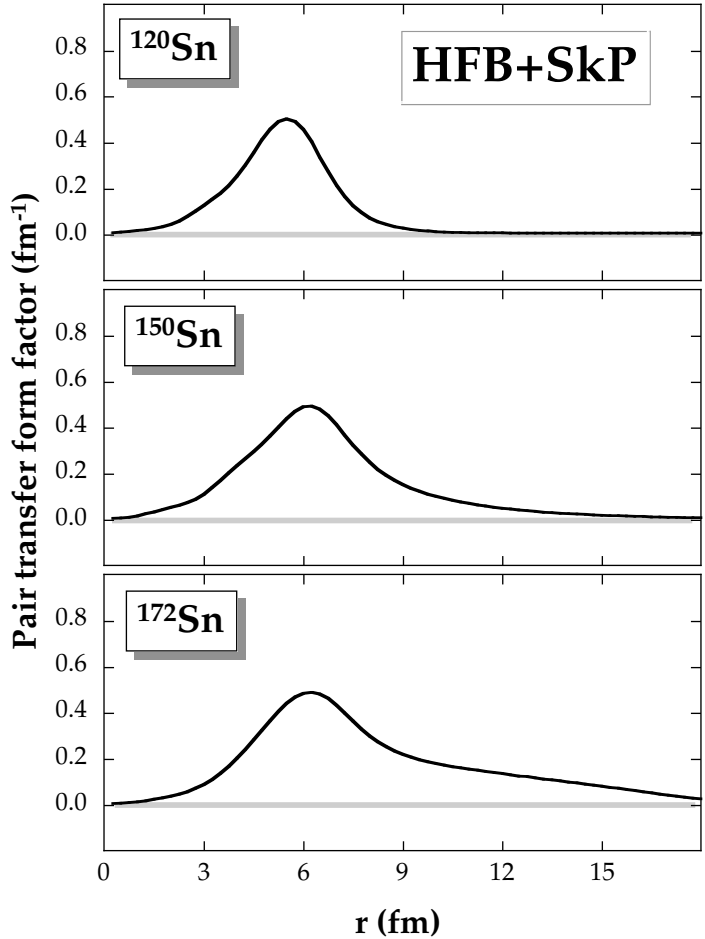}}
\vspace*{-0.5cm}
\caption{
Pair transfer form factor, $r^2\tilde\rho(r)$, calculated directly
from the HFB pairing density $\tilde\rho(r)$.
(From Ref.~\protect\cite{[Dob96]}.)
}
\label{transfer}
\end{figure}

Figure~\ref{transfer} displays the pair transfer form factors
$r^2\tilde\rho(r)$
calculated in $^{120}$Sn, $^{150}$Sn, and $^{172}$Sn
with the SkP interaction.
These pair transfer form factors
clearly show that this process
has a predominantly surface character.
In particular,
there is a
significant increase in  the pair
transfer form factors in the outer regions of drip-line nuclei.
In $^{120}$Sn, the form factors vanish around 9\,fm, while
in $^{150}$Sn and $^{172}$Sn
they extend to much larger distances.

\subsection{Shell structure far from stability and position of the neutron
drip line}\label{shells}

The structure of nuclei
is expected to change significantly as the limit of nuclear
stability is approached   in neutron excess.
Due to the systematic variation in the spatial distribution of nucleonic
densities
 and the increased  importance of the pairing field, the
average nucleonic potential is modified
 when approaching the neutron drip line. The main effect is the increase
of the potential  diffuseness;  the single-particle
neutron potential in drip-line nuclei
becomes very shallow, and  the resulting
shell-model spectrum resembles that of a harmonic oscillator
with a spin-orbit term and with a weakened ${\ell}^2$ term
\cite{[Dob94]}. This results in a
new shell structure characterized by a more
uniform distribution of normal-parity orbits and the
unique-parity intruder orbit which reverts towards its parent
shell.

\begin{figure}[tbh]
\epsfxsize 11.cm
\centerline{\epsfbox{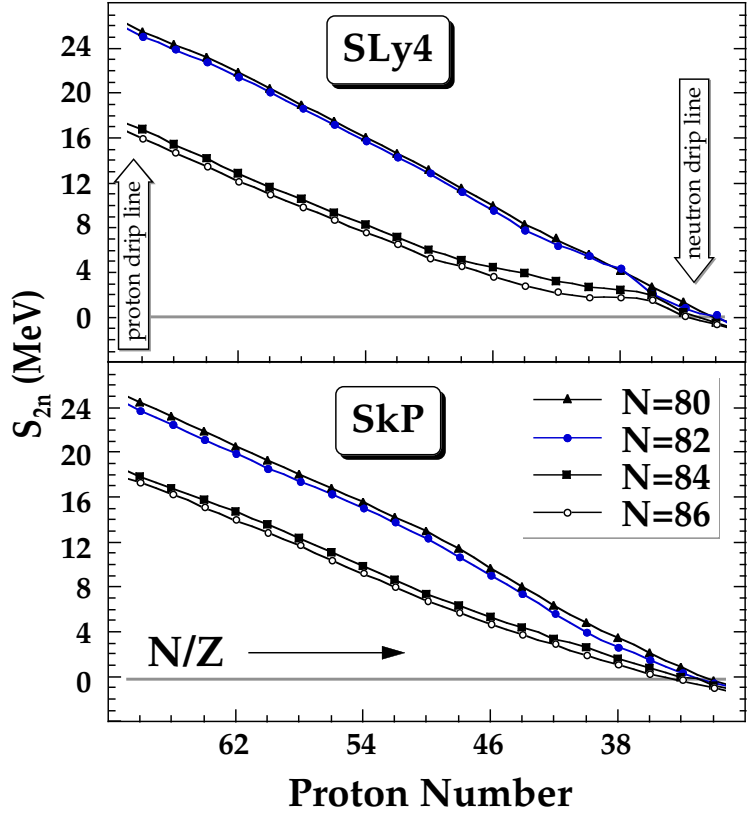}}
\vspace*{-3cm}
\caption{
Two-neutron separation energies for the $N$=80, 82, 84, and 86
spherical even-even isotones calculated in the HFB+SkP
and HFB+SLy4$^\delta$ models as
functions of the proton number.
The arrows indicate the proximity of neutron and
proton drip lines for small and large proton numbers, respectively.
}
\label{gap82}
\end{figure}

The effect of the weakening of shell effects in drip-line
nuclei, first mentioned in Ref.~\cite{[Ton78]},
was further investigated in
Refs.~\cite{[Hae89],[Smo93],[Dob94],[Dob95c],[Che95],[Pea96]}.
Quenching  of shell effects manifests itself
in the behavior of two-neutron separation energies $S_{2n}$.  This is
illustrated in Fig.~\ref{gap82} which displays the two-neutron
separation energies for the $N$=80, 82, 84, and 86 spherical
even-even isotones calculated in the HFB model with
the SkP \cite{[Dob84]} and SLy4 \cite{[Cha95a]} effective interactions.
  The  large $N$=82 magic gap, clearly seen in
the nuclei close to the stability valley and to the proton drip
line, gradually closes down when approaching  the neutron drip
line. This result is independent of the size of the $N$=82 shell
gap in stable nuclei, which is slightly underestimated and
overestimated by SkP and SLy4 forces, respectively, as compared to
experimental data. It can be attributed to two effects: (i) a gradual
increase of the neutron surface diffuseness across the stability
valley related to an increase of the neutron excess, and (ii) the
influence of the continuum, which results in closing the shell gap
near the neutron drip line down to zero.

\begin{figure}[tbh]
\epsfxsize 10.cm
\centerline{\epsfbox{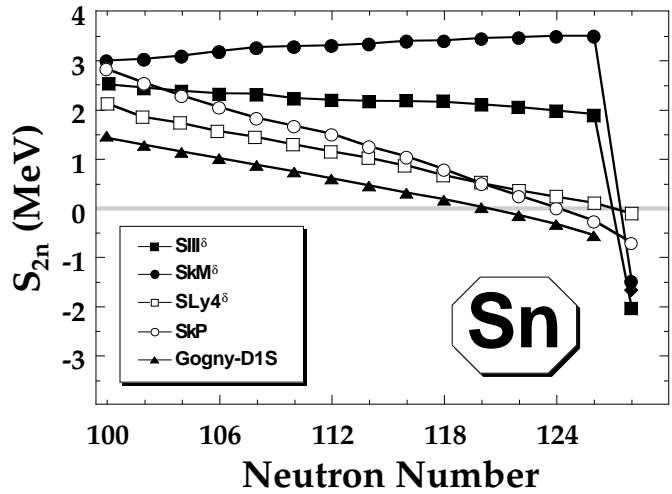}}
\caption{
Two-neutron separation energies $S_{2n}$
for the Sn isotopes,  calculated
in the HFB approach with the Skyrme interactions
SIII$^\delta$, SkM$^\delta$, SLy4$^\delta$, and SkP
and with the
Gogny-D1S interaction.
}
\label{sndrip}
\end{figure}

Predicted behaviour of the two-neutron separation energies
depends very much on the effective interaction used.
This is illustrated in Fig.~\ref{sndrip} which shows
the $S_{2n}$  values calculated in tin isotopes for the Gogny
interaction D1S and for four variants of the Skyrme interaction.
The Gogny force and the SkP and SLy4$^\delta$ Skyrme forces
predict a gradual decrease of the two-neutron separation energies
while the older Skyrme forces, SIII$^\delta$ and SkM$^\delta$,
give almost constant values followed by a sudden drop at $N$=126.
This shows that the $N$=126 shell quenching is not a generic effect.

As seen in Fig.~\ref{sndrip}, the position of the neutron
drip line for the Sn isotopes also depends on the effective interaction used;
it varies between  $N$=118 (D1S) and  $N$=126 (SIII$^{\delta}$).
Hence, even
if the theoretical method used to calculate nuclear masses
is reliable near the drip line (this is not the case for the commonly used
macroscopic-microscopic models, see Ref.~\cite{[Naz94]}),
 the uncertainty due to the largely unknown
isospin dependence of the force gives an
appreciable  theoretical ``error bar".

\begin{figure}[htb]
\epsfxsize 16.cm
\centerline{\epsfbox{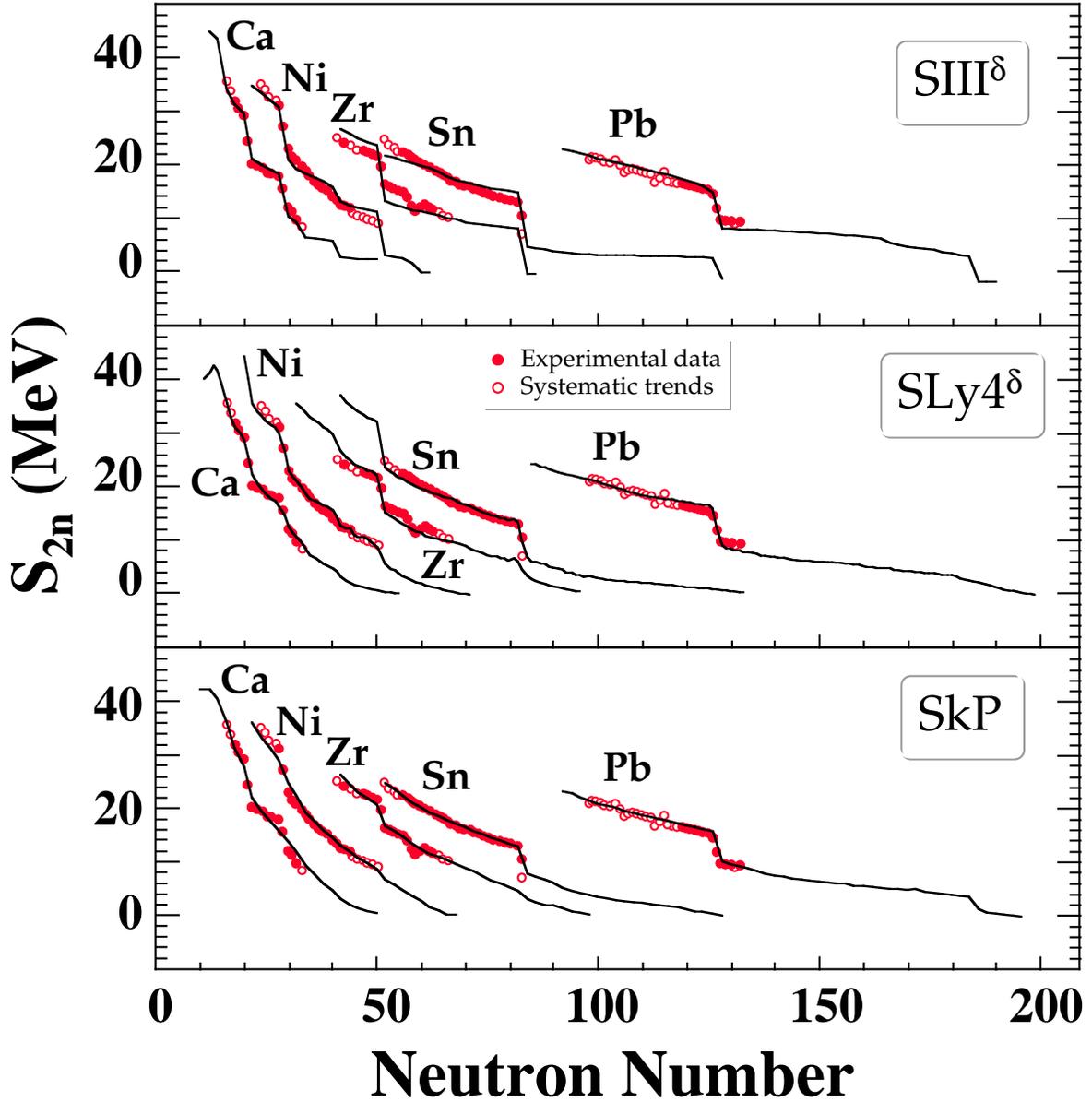}}
\vspace*{-5cm}
\caption{
Two-neutron separation energies, $S_{2n}$, for
             proton-magic isotopes. The HFB results with
             the SIII$^\delta$, SLy4$^\delta$, and
             SkP parametrizations  (solid lines)
             are compared with the experimental data (full circles)
             and systematic trends (open circles)
\protect\cite{[Aud93]}.
}
\label{S2n}
In the following section we discuss another important aspect of
the shell quenching, namely, its consequences for the r-process
and the stellar nucleosynthesis.

\end{figure}

Unfortunately, the results presented in
Fig.~\ref{sndrip} do not tell us much about which of the
forces discussed should be the ``preferred one" since we are
dealing with dramatic extrapolations far beyond the region known
experimentally.
The comparison with the data can be carried out for nuclei
closer to stability, see Ref.\ \cite{[Pat96]} for a recent
quantitative analysis.
A comparison with the experimental two-neutron
separation energies is displayed in
Fig.~\ref{S2n}. As seen,  the agreement with data
is unsatisfactory
for SIII$^\delta$.
In particular, the shell-gap sizes at $N$=50, 82, and 126
are strongly overestimated, and
 the values and the slopes of $S_{2n}$
are, in most cases, incorrect. For the SkM$^\delta$ force
one obtains similar deficiencies \cite{[Dob95c]}.
The results shown in Figs.~\ref{sndrip} and \ref{S2n}
illustrate a very strong dependence of the
two-neutron separation energies on the force parameters.
Although older forces, such as SIII and SkM*,
can perform well in certain regions of $Z$ and $N$, they
do not give a satisfactory global reproduction
of the data. On the other hand,
a fairly good global agreement obtained with  the SkP and SLy4$^\delta$
suggests that the
improvement is possible, while a still better parametrization
would be welcome.
Of course, forces which fail in reproducing the
behavior with ($N-Z$) in known nuclei have little chance to perform
better
when going far from stability. For example, the predicted
values of $S_{2n}$ obtained for the SIII$^\delta$ and  SkM$^\delta$
interactions (Fig.~\ref{sndrip}) followed by a strong
shell effect at $N$=126 do not seem
very reliable.
A  detailed analysis of the force-dependence
of results  may give us
valuable
information on the relative importance of various force parameters.

The gradual change in  shell structure is expected to
give rise to new sorts of collective  phenomena
\cite{[Naz94],[Cho95]}. It is also to be noted that the
experimentally observed collapse of magic gaps seen in some neutron-rich
light nuclei is conventionally explained in terms of the
shape
transition to the deformed intruder configuration.
Here, spectacular examples
are $^{32}$Mg$_{20}$ \cite{[Det79],[Tou82]} and $^{44}$S$_{28}$
\cite{[Sch96],[Gla97]}. In both cases, HF calculations  predict the
 shape transition  \cite{[Cam75],[Wer94a]} due to the  crossing of spherical
configuration
by intruder states.
Such a lowering of the intruder configuration
depends on the detailed balance between three components in
the total energy: (i) the
position of the intruder state at the spherical shape,
(ii) the deformation energy
gain associated with the deformation-driving orbital,
and (iii) the symmetry-restoring force exerted by particles in the
magic-shell configuration. The first and the third of these
elements directly depend
on the spherical shell structure and thus can be affected by the hypothetical
shell quenching mechanism. However, the appearance of a coexisting
configuration cannot, of course, be discussed solely in terms of the
spherical shell structure, and it still remains
an open
and exciting problem.

\subsection{Structure of neutron-rich nuclei and the r-process}\label{astro}

The very neutron-rich drip-line nuclei
cannot be reached experimentally under present
laboratory conditions.  On the other hand, these systems are the
building blocks of the astrophysical r-process; their separation
energies, decay rates, and neutron capture cross sections are the basic
quantities determining the results of nuclear reaction network
calculations.
 Consequently, one hopes to learn  about properties of
very neutron-rich systems by studying the r-process component
of the solar-system abundances of heavy elements
\cite{[Kra93],[How93],[Pea96],[Pfe97],[Sur97]}.  The recent r-process
 network calculations
\cite{[Kra93],[Che95],[Pfe97]}, based on several mass formulae, indicate
that a quenching of the  shell effect at $N$=82
is required in order to fill the
$A$=120 and 140 r-abundance troughs,
in  accordance with the
results of the HFB+SkP model shown in Fig.~\ref{gap82}.

 In addition to nuclear structure,
there are other factors which can influence the  r-process abundances,
for instance the astrophysical conditions of
temperature, neutron density, and the process time scale \cite{[Gor96]}.
The possibility that abundances of r-process elements may
be altered by the intense
neutrino flux has been discussed in Refs.~\cite{[Hax97],[Qia97]}.
According to their calculations,  neutrino reactions
can be important in breaking through the waiting-point nuclei
at $N$=50 and 82,
and  the r-process abundances in the
$A$=125 and 185 regions
can be affected by neutrino post-processing effects.

In addition to nuclear masses, another important piece of nuclear structure
that determines the path of the r-process is the GT strength.
It is usually calculated in the quasiparticle RPA (QRPA) theory
\cite{[Kra93],[Sta89],[Mol90a],[Tac95]}. The important development,
yet to be done, would consist of performing systematic
microscopic QRPA calculations
based on  the HFB densities. This would guarantee the proper treatment
of the particle continuum in the weakly bound nuclei on the
r-process path.

\subsection{Deformation of drip-line nuclei}

Neutron halos and heavy, weakly bound  neutron-rich nuclei
offer an opportunity to study the wealth of phenomena
associated with the closeness of the particle threshold:
particle emission (ionization to the continuum) and
characteristic behavior of cross sections \cite{[Wig48],[Fan61]},
 existence of soft
collective modes  and low-lying transition strength
\cite{[Uch85],[Fay91],[Yok95],[Sag95],[Ham96]},
as well as
various other nuclear properties in the sub-threshold regime.
We have learned that  weakly bound nuclei
are different; they have giant sizes, they are diffused,
they are strongly superfluid, their shell structure is probably different.
But can they be deformed?

 The importance of non-spherical
intrinsic shapes in halo nuclei  has been discussed,
especially in the context of a one-neutron halo in $^{11}$Be.
The ground state of  $^{11}$Be is a 1/2$^+$ state. The low neutron
separation energy, $S_n$=504 keV, allows for only one bound excited level
(1/2$^-$ at 320 keV). The halo character of $^{11}$Be
has been confirmed by studies of reaction cross sections \cite{[Fuk91a]},
and the importance of  deformation can be inferred from the large
quadrupole moment of its  core
$^{10}$Be, $|Q|$=229\,mb  \cite{[Ram87]}.
 The role of deformation in lowering the excitation energy of the  1/2$^+$
intruder  level  in $^{11}$Be has been recognized
 \cite{[Bou78],[Rag81]}, but the joint  effect
of  loose binding
and deformation has not been considered. (See, however, recent
references \cite{[Kan95],[Esb95],[Vin95],[Ber95],[Li96]}.)

In a recent  study \cite{[Mis97]},
  the notion of shape deformations in
halo nuclei has been addressed
 by considering
the single-particle motion in
 the axial spheroidal square well.
The properties of the deformed single-particle
states, especially in the subthreshold region, were analyzed by making
the multipole decomposition in the spherical partial waves
with well-defined
orbital angular momentum.
It has been concluded that
in the
limit of  very weak binding,
the geometric  interpretation of shape deformation is lost.
That is, the deformation of the halo is solely determined by
the spatial structure of the valence state wave function,
independently of the shape of the core. The deformed
core merely establishes  the quantization axis of the system
-- important for determining the angular momentum projection
on the symmetry axis, $\Lambda$.

\begin{figure}[tbh]
\epsfxsize 15.cm
\centerline{\epsfbox{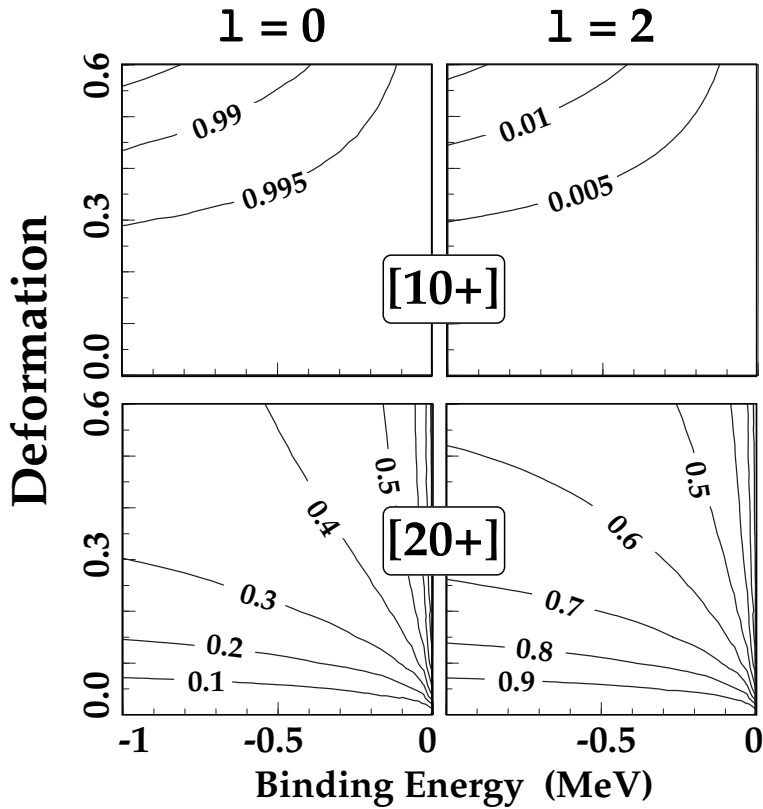}}
\caption{
Contour maps of probabilities $P_0$ and $P_2$
 for the [10+] (top) and [20+] (bottom) Nilsson
levels as functions of deformation and binding energy.
 (From Ref.~\protect\cite{[Mis97]}.)
}
\label{ppp}
\end{figure}

Figure~\ref{ppp}
shows the contour map of $P_{\ell}$ (probability to find the partial wave
$\ell$ in a given Nilsson state [$n_{\rm exc}\Lambda\pi$])
for the $\Lambda$=0 orbitals
as functions of binding energy  and deformation.
The structure of the [10+]
Nilsson level, originating from the
spherical $1s$ state,
  is completely dominated by the $\ell$=0 component, even at very
large deformations. A rather interesting pattern is seen in the
diagram for the [20+] orbital
originating from the
spherical $1d$ state. The $\ell$=2 component dominates at low and medium
deformations, and the corresponding probability
$P_{\ell=2}$  slowly
decreases with $\delta$  at large deformations approaching
the (constant) asymptotic limit.
 However, a similar effect, namely
the decrease of the  $\ell$=2
component, is seen when approaching the zero binding energy  threshold.
In the language of the perturbation theory \cite{[Fan61]},
 this rapid transition
comes from
the coupling to the low-energy $\ell$=0 continuum.
As a consequence of the dominating role of the $s$-wave
(and $p$-wave, for negative-parity states),
in the limit of weak binding,
the total quadrupole deformation of the (core+valence) system
depends solely on the geometry of the valence orbital. Namely,
it is consistent with
a superdeformed shape  ($\pi$=--, $\Lambda$=0 halo), a
spherical shape  ($\pi$=+, $\Lambda$=0 halo), or an
oblate shape ($\pi$=--, $\Lambda$=1 halo), regardless of the deformation of
the core.
In the language of the self-consistent mean-field theory, this result
reflects the extreme softness of the system to the quadrupole distortion.
Shape
deformation is an extremely  powerful concept provided that the
nuclear surface can be properly defined. However,
 for very diffused and spatially extended systems,
the geometric interpretation of multipole moments and deformations
is lost.

The presence of the spatially extended  neutron halo
gives rise to  the  low-energy isovector modes.
The deformation decoupling of the halo
implies that the nuclei close to the neutron drip line are
excellent candidates for isovector quadrupole deformations,
 with different
quadrupole deformations for protons and neutrons.
An example of such a situation has been predicted in the self-consistent
calculations
for the neutron-rich sulfur isotopes
performed
 using the Skyrme
HF  and  relativistic mean-field
methods \cite{[Wer94a],[Wer96]}.
When approaching
the neutron drip line, the calculated values of
quadrupole deformation
for neutrons are systematically
smaller than those
 of the proton distribution.
Another promising candidate for such effects is the $^{120}$Sr nucleus
which is presently being studied by the GCM configuration mixing of
the Skyrme HF states \cite{[Nau96]}.

The discussion of deformation in halo nuclei applies to systems with
very small binding energy and with negligible pairing. How will this scenario
be modified in the presence of pairing, and for greater separation energies?
We do not know the answer to this question at present;
the systematic investigation of the interplay between deformation
and pairing in weakly bound neutron-rich nuclei
is one of the most exciting avenues of
RNB physics.

\section{Physics of Proton-Rich Nuclei: at and  Below
the  $N/Z$=1 Limit}\label{smallNZ}

On the proton-rich side of the valley of stability,
physics is different than in nuclei
with a large neutron excess.
Because of the Coulomb barrier which tends to localize the proton
density in the nuclear interior,
nuclei beyond the proton drip line are quasibound
with respect to proton emission. However, in spite of the stabilizing
effect of the Coulomb barrier, the effects associated with the weak binding
are also present in proton drip-line nuclei. They are not as dramatic
as on the other side of the stability valley, but nevertheless important.
For instance, the Thomas-Ehrman
 shift  \cite{[Tho51],[Ehr51]}, which is due to changes in the Coulomb energy
of the weakly bound proton, can lead to a
decrease in the energy differences between analog states by a few hundred keV.
This effect is most significant for loosely bound states
 and for orbitals  having
low angular momentum \cite{[Sch78a]}. (For the self-consistent calculations
of the Thomas-Ehrman
 shift in the doubly-magic diproton emitter
$^{48}$Ni, see Ref.~\cite{[Naz96]}.)
Consequently, indiscriminate applications of the nuclear shell-model
to nuclei
close to and beyond the proton drip line, ignoring the systematic
changes in single-particle energies  and wave functions due to
weak binding, should probably be taken with a grain of salt.

The doubly magic $N$=$Z$=50 nucleus $^{100}$Sn
is a   paradigm of RNB physics at the
 proton-rich side. Although it was found
experimentally three years ago
\cite{[Sch94],[Lew94]}, it took more than two years to
roughly determine its mass \cite{[Cha96]}, and it will probably take
quite a few years
to find its first excited state. Actually, the question what is this
state constitutes an unresolved problem which is
a challenge for theoretical predictions.

\subsection{Neutron-proton correlations}

 A unique aspect of proton-rich
nuclei with  $N$=$Z$ is that neutrons and protons occupy the same
shell-model orbitals. Consequently,
due to
the large spatial overlaps between neutron
and proton single-particle wave functions,
the proton-rich $N$=$Z$ nuclei are expected to exhibit unique manifestations
of  proton-neutron (pn) pairing \cite{[Gos64],[Gos65],[Cam65],[Bar69],[Che78]}.

At present, it is not clear what the specific experimental
fingerprints of the pn  pairing are, whether the
pn  correlations
are strong enough to form
a static pair condensate, and what are their
main building blocks \cite{[NP97]}.
Most of our knowledge about nuclear pairing
 comes from nuclei with a  sizable neutron excess where
the isospin $T$=1 neutron-neutron  and proton-proton
 pairing  dominate. Now, for the first time,
there is an experimental opportunity to explore
nuclear  systems in the
vicinity of the $N$=$Z$ line which have
 {\em many} valence $np$ pairs;
that is,  to probe the interplay between the like-particle and
pn  ($T$=0,1) pairing channels.

This novel situation calls for  the generalization of
established theoretical models  of nuclear pairing.  In spite of
several early attempts  to extend  the
independent quasi-particle formalism to incorporate the effect
of pn  correlations in light nuclei
 (see Ref.~\cite{[Goo79]} for
an early
review), no symmetry-unrestricted calculations  for np
pairing, based on the isospin-projected quasi-particle theory,
have been carried out.

So far, the strongest evidence for enhanced pn correlations
around the $N$=$Z$ line  comes from the
measured binding energies. An additional binding (the so-called
Wigner energy) found in these nuclei  manifests itself as a
spike in the isobaric mass parabola as a function of
$T_z$=$\frac{1}{2}(N-Z)$ (see
the review \cite{[Zel96]} and Refs.  quoted therein).
The  pn  correlations are
 also  expected to
play a role in beta decay \cite{[Che93],[Che93a],[Sch96a],[Eng97]},
deuteron transfer reactions \cite{[Fro70],[Fro71]},
structure of high spins \cite{[Goo79],[Kva90],[Sat97a]},
and also in nuclear matter
\cite{[Pal75],[Alm90],[Von91]}.

\begin{figure}[tbh]
\epsfxsize 15.cm
\centerline{\epsfbox{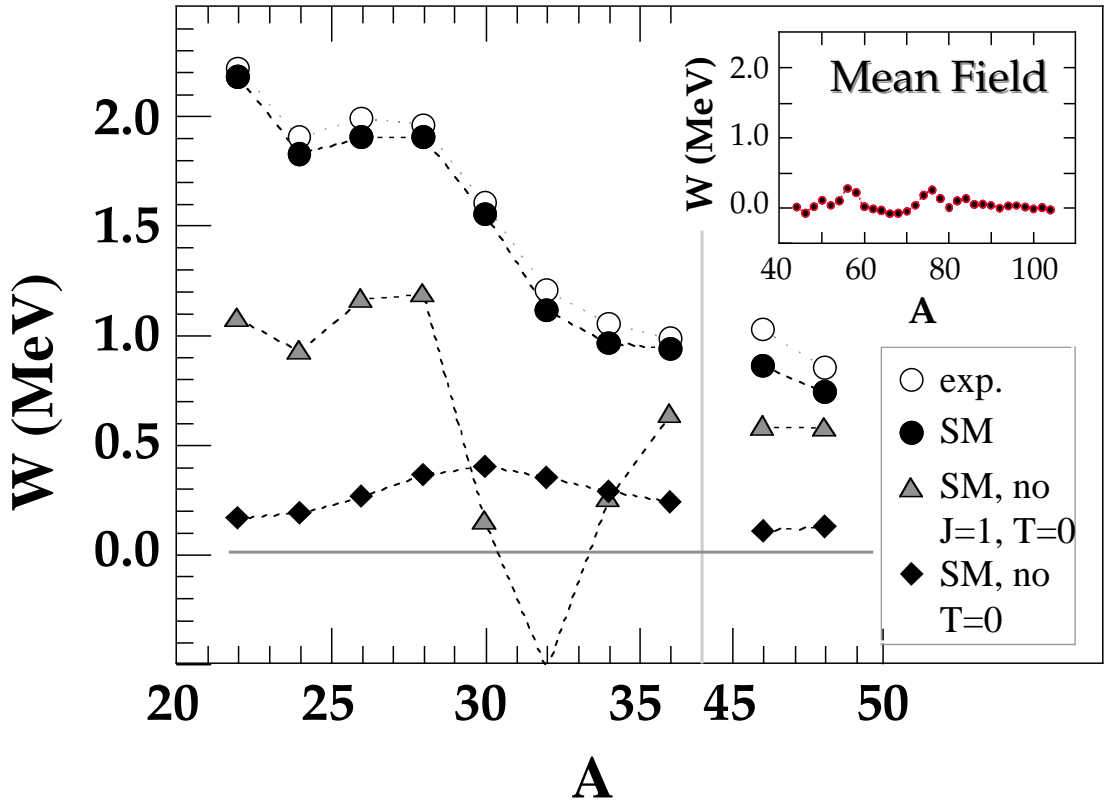}}
\caption{
The strength of
the Wigner term, $W$,
 extracted using binding energies calculated with the
$0\hbar\omega$
shell model.  Full shell-model calculations
(filled circles)
 agree very well with experimental data (open circles).
The  results of shell-model
calculations with the
($J$=1, $T$=0) two-body matrix elements removed
($J_{\rm max}$=1 variant, triangles) and
with all  $T$=0 matrix elements removed
($J_{\rm max}$=7 variant, diamonds),
are also shown.
The inset shows the  values of $W$
extracted from the ETFSI mass formula
\protect\cite{[Abo95]}. They are practically zero for all nuclei
considered. (From Ref.~\protect\cite{[Sat97]}.)
}
\label{wigner}
\end{figure}

The role of the $T$=0 part of the interaction
on the presence of binding-energy
irregularities
near the  $N$=$Z$ has been recognized
in Ref.~\cite{[Bre90]}.
Recent calculations~\cite{[Sat97]}
have revealed the rather complex mechanism
responsible for the
nuclear binding around the $N$=$Z$ line.
In particular,
it has been  found that the Wigner term cannot be
solely explained in terms of correlations
between  the proton-neutron
$J$=1, $T$=0  (deuteron-like) pairs (see Fig.~\ref{wigner}).
(For more discussion of this point, see also
Ref.~\cite{[Eng96]}.)

Recently, the isospin structure of the
density matrices and
self-consistent mean fields  has been discussed
\cite{[Per96]} in the  HFB
theory allowing for a consistent microscopic description of
pairing correlations in all isospin channels.
Theoretically, the pn pairing correlations have been
studied by several authors in the HFB framework
\cite{[Goo72],[Wol71]}.
However, in Ref.~\cite{[Per96]} this has been
done  in the
coordinate space allowing for the
classification of generic proton-neutron mixing mean fields.
The resulting HFB equations have interesting properties.
For spherical nuclei, only the  $T$=1 and $J$=0 nucleonic pairs
are allowed. (The presence of the  $T$=0 and $J$$\ne$0 pairs  would necessarily
lead to deformed mean fields.)
Consequently, for the spherical symmetry, two cases can be considered.
The first one corresponds to pn pairs coupled
to $S$=0 and is similar to the standard like-particle pairing.
The second one, analogous to triplet pairing with
  $S=1$, is  more interesting
because of breaking of the intrinsic parity.
Here,
the HFB solution  contains mixtures
of proton and neutron states with different parities
but with the same value of $j$,  e.g.,
a mixing of the $\nu g_{7/2}$ and $\pi f_{7/2}$ orbitals.

\subsection{Proton emitters}

Nuclei beyond the proton drip line
are ground-state proton emitters.
Initially, the  parent nucleus
is in a quasistationary state, and the proton emission
may be considered as a process where the proton tunnels through
the potential barrier.  In most cases, the combined
Coulomb and centrifugal potentials
give rise to barriers which are
as large as $\sim$15\,MeV. Consequently,
the  associated lifetimes, ranging from 10$^{-6}$\,sec to a
few seconds, are sufficiently long to obtain a wealth of spectroscopic
information. Experimentally, a number of  proton emitters have now been
discovered in the mass regions $A$$\sim$110, 150, and 170
(see Refs.~\cite{[Hof95b],[Hof97]} and
references
quoted therein).
It is anticipated that new regions of proton-unstable nuclei will be
explored in the near future using radioactive nuclear beams.

The width
of the proton resonance  can be estimated
through  the distorted-wave Born approximation (DWBA) \cite{[Kad71]}.
From the decay width one can obtain the
half-life of the proton emission,
$t_{1/2}$.
The proton resonances are extremely
narrow, $\Gamma \sim 10^{-22} - 10^{-15}$\.MeV, hence it is
difficult to
calculate their widths directly. (It is worth noting that
in Ref.~\cite{[Fei83]} an attempt was made
to calculate the proton emission width by solving
the Schr\"odinger equation in the complex plane.)

Recently, half-lives of the spherical proton emitters have
been calculated in Ref.~\cite{[Abe97]}
in the core-plus-proton approach assuming $V_{Ap}$
to be  a sum
of a simple nuclear optical Woods-Saxon potential
and the Coulomb potential.
Three different methods have been used: DWBA,
the modified  two-potential approach of Gurvitz \cite{[Gur88]},
 and the
semi-classical approximation WKB.
(Because of its simplicity, the
 WKB  approach has been widely used to study spherical
proton emitters \cite{[Buc92]}.)

After computing
the  barrier
penetration factor,
the experimental spectroscopic factors $S_p^{\rm exp}$
could be determined as
 ratios of
calculated and measured half-lives \cite{[Hof97]}.
 Theoretically, the spectroscopic factor
measures the fragmentation of a single-particle orbital $(n\ell j)$.
In the BCS
theory, it  is given by
$u_j^2$, i.e.,  the  probability
that  the spherical orbital ($n \ell j$) is empty
in the daughter nucleus.

\begin{figure}[tbh]
\epsfxsize 15.cm
\centerline{\epsfbox{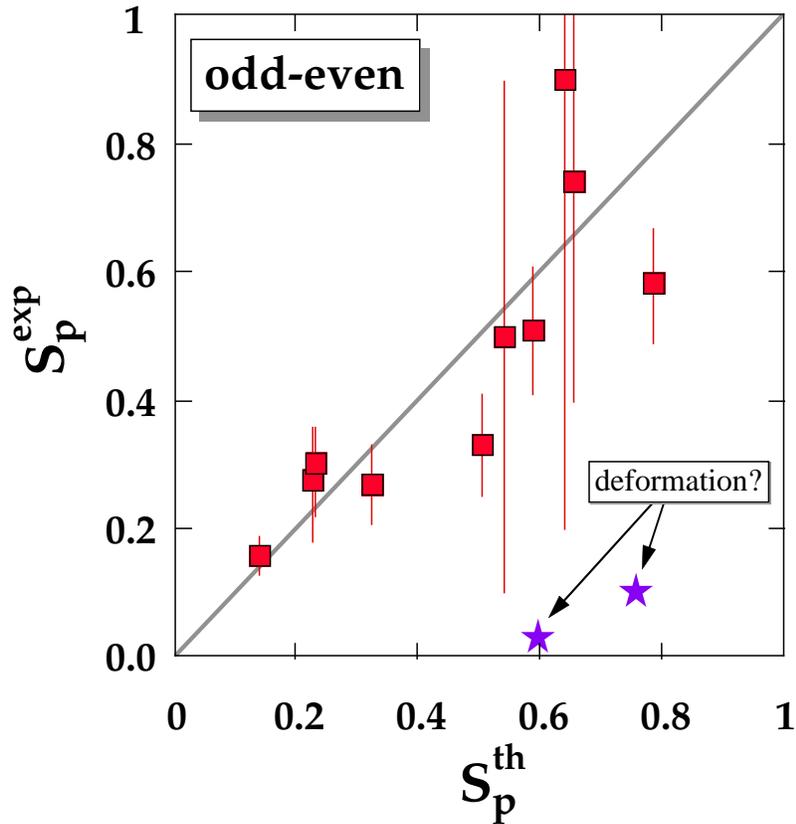}}
\caption{
Correlation between the experimental
proton spectroscopic factors
$S^{\rm exp}_p$ deduced from measured ground-state
proton emission half-lives
of odd-$Z$, even-$N$ proton emitters
 and  theoretical
values $S^{\rm th}_p$ obtained in the BCS theory.
The nuclei expected to be deformed are
indicated by stars.  (From Ref.~\protect\cite{[Abe97]}.)
}
\label{Sfact}
\end{figure}
The correlation between experimental and theoretical
spectroscopic factors obtained in the DWBA calculations
is shown in Fig.~\ref{Sfact} for
 odd-$Z$, even-$N$ ground-state proton emitters.  The agreement
between experiment and theory is good. For the two cases indicated by stars,
$^{109}$I and $^{113}$Cs, the experimental values fall well below
theoretical predictions. This suggests a strong fragmentation of the
single-particle strength and/or increased tunneling probability
as compared to spherical predictions.  Indeed, both
 $^{109}$I and $^{113}$Cs are predicted to be deformed, and
the ``anomalous" proton half-life of $^{109}$I has been reproduced
by deformed calculations of Ref.~\cite{[Kad96]}.

In general,
proton emission half-lives depend mainly on the proton
separation energy and orbital angular momentum,
but rather weakly on the details of intrinsic structure
of proton emitters, e.g., on
the parameters of the
proton potential
at least at a qualitative level (factors of 2-3).
The weak sensitivity of $t_{1/2}$ to
the details of the optical proton potential
 has been discussed in Ref.~\cite{[Naz96]}
in the context of two-proton radioactivity.
It has been shown that  more than 94-99\% of
the WKB exponent
 comes from the region $r$$>$$r_B$, which is almost
solely determined by the combined Coulomb and centrifugal potentials.
 This suggests that the lifetimes of deformed proton
 emitters will provide  direct information on the angular momentum
content of the associated Nilsson state, and hence, indirectly
on the nuclear shape.

Proton radioactivity is an
excellent example of the elementary three-dimensional
quantum-mechanical tunneling. Experimental and theoretical
investigations of proton emitters  (or theoretically predicted
ground-state di-proton emitters) will open up a  wealth of
exciting physics associated with the
residual interaction coupling between bound states and extremely
narrow resonances  in  the region of very low density of
single-particle levels.

\section{Conclusions}\label{conclusions}

An experimental excursion into new territories
of the chart of the nuclides
will offer
many excellent
opportunities for traditional nuclear structure.
This may include new regions
of quadrupole and octupole deformation, new regions of shape isomers,
including superdeformations, new combinations of magic or semi-magic
closures, and many others.
For a comprehensive review of these possibilities, the reader is referred to
Ref.~\cite{[ISL91]}.

In trying to see the phenomena of a
``new physics", we should ask the fundamental question of
``how far is far"?
Experiments with radioactive beams
are going to be long and difficult, and many examples of nuclear exotica
discussed in this paper (especially those concerning
very neutron-rich systems) are clearly out of reach,
even assuming most optimistic experimental
scenarios. The hope is, however, that
some of the effects associated with the loose binding will be seen
as deviations from smooth systematic trends \cite{[Cho95]}
or will show up  at higher excitation energies
closer to the particle threshold, as in the example of the analog states
\cite{[Tho51],[Ehr51]}.
Theoretically, we are bound to adopt the strategy of going to
the extreme values of $N$/$Z$ in order to identify the qualitatively
new phenomena, and then back down to experimentally achievable
regions to see whether these phenomena can actually be observed.
There is very little  doubt that
we are on the verge of the most fascinating fishing expedition;
a lot of  exciting physics
will probably be caught already  at the beginning of
this journey.

The main objective of this study
was to discuss
various theoretical facets of nuclear structure with radioactive beams.
In
particular,  the unusual conditions created by the weak binding
and the importance of the
coupling to the particle continuum have been emphasized.
The theoretical formalism has been applied
to experimental observables; i.e.,  energy spectra,
masses, radii, surface thickness, and pair transfer form factors. It is
demonstrated that these observables carry invaluable information
that can pin down many basic questions regarding the effective
nucleon-nucleon interaction.

The analysis presented in this paper should be viewed as a
useful starting point for future investigations. One of them is
the coupling between vibrational and rotational  modes
 and pairing fields in weakly bound nuclei. Another
interesting avenue of exploration is the role of dynamics;
e.g., the importance of the particle number conservation and
isospin mixing.
 A fascinating and difficult
research program is the microscopic description of excited
states, especially those lying above the particle emission
threshold.
We are only beginning to explore many
unusual  aspects of the nuclear many-body problem offered by
systems with extreme $N/Z$ ratios.

\acknowledgments

The Joint Institute for Heavy Ion
 Research has as member institutions the University of Tennessee,
Vanderbilt University, and the Oak Ridge National Laboratory; it
is supported by the members and by the Department of Energy
through Contract No. DE-FG05-87ER40361 with the University
of Tennessee.
This research
 was supported in part by the U.S. Department of
Energy through Contract No. DE-FG02-96ER40963 with
the University
of Tennessee, and
by the Polish Committee for
Scientific Research under Contract No.~2~P03B~034~08
with the Warsaw University.
Oak Ridge National
Laboratory is managed for the U.S. Department of Energy
 by Lockheed Martin Energy Research
Corporation  under Contract No.
DE-AC05-96OR22464.


\begin{thebibliography}{100}

\bibitem{[Dab77]}
{J. D{\c a}browski, Nukleonika {\bf 22}, 143 (1977)}.

\bibitem{[Fre86]}
{J. Freidrich and P.G. Reinhard, Phys. Rev. {\bf C33}, 335 (1986)}.

\bibitem{[Pea94]}
{J.M. Pearson and M. Farine, Phys. Rev. {\bf C50}, 185 (1994)}.

\bibitem{[Sha95]}
{M.M. Sharma, G. Lalazissis, G. K\"onig, and P. Ring, Phys. Rev. Lett. {\bf
  74}, 3744 (1995)}.

\bibitem{[Rei94]}
{P.-G. Reinhard and H. Flocard, Nucl. Phys. {\bf A584}, 467 (1995)}.

\bibitem{[Cha95]}
{E. Chabanat, P. Bonche, P. Haensel, J. Meyer, and F. Schaeffer, Physica
  Scripta {\bf T56}, 231 (1995)}.

\bibitem{[Ons97]}
{M. Onsi, R.C. Nayak, J.M. Pearson, H. Freyer, W. Stocker, Phys. Rev. {\bf
  C55}, 3166 (1997)}.

\bibitem{[Pud96]}
{B.S. Pudliner, A. Smerzi, J. Carlson, V.R. Pandharipande, S.C. Pieper, and
  D.G. Ravenhall, Phys. Rev. Lett. {\bf 76}, 2416 (1996)}.

\bibitem{[Ber91]}
{G.F. Bertsch and H. Esbensen, Ann. Phys. (N.Y.) {\bf 209}, 327 (1991)}.

\bibitem{[Dob96]}
{J. Dobaczewski, W. Nazarewicz, T.R. Werner, J.-F. Berger, C.R. Chinn, and J.
  Decharg\'e, Phys. Rev. {\bf C53}, 2809 (1996)}.

\bibitem{[Fay96]}
{S.A. Fayans and D. Zawischa, Phys. Lett. {\bf 383B}, 19 (1996)}.

\bibitem{[She96]}
{Y. Shen and Z. Ren, Z. Phys. {\bf A356}, 133 (1996)}.

\bibitem{[Dob95c]}
{J. Dobaczewski, W. Nazarewicz, and T.R. Werner, Physica Scripta {\bf T56}, 15
  (1995)}.

\bibitem{[Kuo97]}
{T.T. Kuo, F. Krmpoti\'c, and Y. Tzeng, Phys. Rev. Lett. {\bf 78}, 2708
  (1997)}.

\bibitem{[Cug87]}
{J. Cugnon, P. Deneye and A. Lejeune, Z. Phys. {\bf A328}, 409 (1987)}.

\bibitem{[Bro88a]}
{G.E. Brown, Phys. Rep. {\bf 163}, 167 (1988)}.

\bibitem{[Gmu92]}
{S. Gmuca, Z. Phys. {\bf A342}, 387 (1992)}.

\bibitem{[Pet95]}
{C.J. Pethick and D.G. Ravenhall, Annu. Rev. Nucl. Part. Sci. {\bf 45}, 429
  (1995)}.

\bibitem{[Pet95a]}
{C.J. Pethick, D.G. Ravenhall, and C.P. Lorenz, Nucl. Phys. {\bf A584}, 675
  (1995)}.

\bibitem{[Dob84]}
{J. Dobaczewski, H. Flocard and J. Treiner, Nucl. Phys. {\bf A422}, 103
  (1984)}.

\bibitem{[Fan61]}
{U. Fano, Phys. Rev. {\bf 124}, 1866 (1961)}.

\bibitem{[Glo67]}
{W. Gl\"ockle, J. Hufner, and H.A. Weidenmueller, Nucl. Phys. {\bf A90}, 481
  (1967)}.

\bibitem{[Iba70]}
{R.H. Ibarra and B.F. Bayman, Phys. Rev. {\bf C1}, 1786 (1970)}.

\bibitem{[Phi77a]}
{R.J. Philpott, Fizika {\bf 9}, suppl. 3, 21 (1977)}.

\bibitem{[Bar77]}
{H.W. Barz, I. Rotter, and J. H\"ohn, Nucl. Phys. {\bf A275}, 111 (1977)}.

\bibitem{[Mic78]}
{M. Micklinghoff, Nucl. Phys. {\bf A295}, 228 (1978)}.

\bibitem{[Hal80]}
{D. Halderson and R.J. Philpott, Nucl. Phys. {\bf A345}, 141 (1980)}.

\bibitem{[Isk91]}
{W. Iskra and I. Rotter, Phys. Rev. {\bf C44}, 721 (1991)}.

\bibitem{[Wen87]}
{W.M. Wendler, Nucl. Phys. {\bf A472}, 26 (1987)}.

\bibitem{[Bol72]}
{M. Bolsterli, E.O. Fiset, J.R. Nix, and J.L. Norton, Phys. Rev. {\bf C5}, 1050
  (1972)}.

\bibitem{[Naz94]}
{W. Nazarewicz, T.R. Werner, and J. Dobaczewski, Phys. Rev. {\bf C50}, 2860
  (1994)}.

\bibitem{[Ben96]}
{J.R. Bennett, J. Engel, and S. Pittel, Phys. Lett. {\bf B368}, 7 (1996)}.

\bibitem{[Ghi96]}
{F. Ghielmetti, G. Colo, E. Vigezzi, P.F. Bortignon, and R.A. Broglia, Phys.
  Rev. {\bf C54}, R2143 (1996)}.

\bibitem{[Vaa79]}
{J.S. Vaagen, B.S. Nilsson, J. Bang, and R.M. Ibarra, Nucl. Phys. {\bf A319},
  143 (1979)}.

\bibitem{[Raw82]}
{G. Rawitscher, Phys. Rev. {\bf C25}, 2196 (1982)}.

\bibitem{[Bub91]}
{M. Buballa, S. Dr\'o\.zd\.z, S. Krewald, and J. Speth, Ann. Phys. {\bf 208},
  346 (1991)}.

\bibitem{[Rid97]}
{D. Ridikas, M.H. Smedberg, J.S. Vaagen, and M.V. Zhukov, Europhys. Lett. {\bf
  37}, 385 (1997)}.

\bibitem{[Rom72]}
{W.J. Romo, Nucl. Phys. {\bf A191}, 65 (1972)}.

\bibitem{[Ber68]}
{T. Berggren, Nucl. Phys. {\bf A109}, 265 (1968)}.

\bibitem{[Ver87]}
{T. Vertse, P. Curutchet, and R.J. Liotta, Lecture Notes in Physics {\bf 325}
  (Springer Verlag, Berlin 1987), p. 179}.

\bibitem{[Lin94]}
{P. Lind, R.J. Liotta, E. Maglione, and T. Vertse, Z. Phys. {\bf A347}, 231
  (1994)}.

\bibitem{[Ber96]}
{J.-F. Berger, L. Bitaud, J. Decharg\'e, M. Girod and S. Peru-Desenfants, Proc.
  of Int. Hirschegg Workshop XXIV, {\sl Extremes of Nuclear Structure} edited
  by H. Feldmeier, J. Knoll, and W. N\"orenberg, (GSI, Darmstadt, 1996), p.
  56}.

\bibitem{[For97]}
{S. Fortunato, A. Insolia, R.J. Liotta, and T. Vertse, Phys. Rev. {\bf C54},
  3279 (1997)}.

\bibitem{[San97]}
{N. Sandulescu, R.J. Liotta, and R. Wyss, Phys. Lett. {\bf B394}, 6 (1997)}.

\bibitem{[Dob96a]}
{J. Dobaczewski, W. Nazarewicz, and T.R. Werner, Z. Phys. {\bf A354}, 27
  (1996)}.

\bibitem{[Naz96]}
{W. Nazarewicz, J. Dobaczewski, T.R. Werner, J.A. Maruhn, P.-G. Reinhard, K.
  Rutz, C.R. Chinn, A.S. Umar, and M.R. Strayer, Phys. Rev. {\bf C53}, 740
  (1996)}.

\bibitem{[Ter96]}
{J. Terasaki, P.-H. Heenen, H. Flocard, and P. Bonche, Nucl. Phys. {\bf A600},
  371 (1996)}.

\bibitem{[Muh84]}
{K. M\"uhlhans, K. Neerg{\aa}rd, and U. Mosel, Nucl. Phys. {\bf A420}, 204
  (1984)}.

\bibitem{[Rei97]}
{P.-G. Reinhard, M. Bender, K. Rutz, and J.A. Maruhn, Preprint
  nucl-th/9705054}.

\bibitem{[Kho82]}
{V.A. Khodel' and \'E.E. Sapershte\u{\i}n, Phys. Rep. {\bf 92}, 12 (1982)}.

\bibitem{[Zve84]}
{M.V. Zverev and \'E.E. Sapershte\u{\i}n, Sov. J. Nucl. Phys. {\bf 39}, 878
  (1984)}.

\bibitem{[Zve85]}
{M.V. Zverev and \'E.E. Sapershte\u{\i}n, Sov. J. Nucl. Phys. {\bf 42}, 683
  (1985)}.

\bibitem{[Smi88]}
{A.V. Smirnov, S.V. Tolokonnikov, and S.A. Fayans, Sov. J. Nucl. Phys. {\bf
  48}, 995 (1988)}.

\bibitem{[Zve91]}
{M.V. Zverev and V.E. Starodubsky, Sov. J. Nucl. Phys. {\bf 54}, 410 (1991)}.

\bibitem{[Men96]}
{J. Meng and P. Ring, Phys. Rev. Lett. {\bf 77}, 3963 (1996)}.

\bibitem{[Poe97]}
{W. Poschl, D. Vretenar, P. Ring, Comp. Phys. Commun. {\bf 103}, 217 (1997)}.

\bibitem{[Mig67]}
{A.B. Migdal, {\sl Theory of Finite Fermi Systems and Applications to Atomic
  Nuclei} (Interscience, New York, 1967)}.

\bibitem{[Shl75]}
{S. Shlomo and G. Bertsch, Nucl. Phys. {\bf A243}, 507 (1975)}.

\bibitem{[Kam93]}
{S. Kamerdzhiev, J. Speth, G. Tertychny, and J. Wambach, Z. Phys. {\bf A346},
  253 (1993)}.

\bibitem{[Sag96]}
{H. Sagawa and C.A. Bertulani, Prog. Theor. Phys. (suppl.) {\bf 124}, 143
  (1996)}.

\bibitem{[Ham96]}
{I. Hamamoto, H. Sagawa, and X.Z. Zhang, Phys. Rev. {\bf C53}, 765 (1996)}.

\bibitem{[Ham96a]}
{I. Hamamoto and H. Sagawa, Phys. Rev. {\bf C53}, R1492 (1996)}.

\bibitem{[Bor95]}
{I.N. Borzov, A.A. Fayans, and E.L. Trykov, Nucl. Phys. {\bf A584}, 335
  (1995)}.

\bibitem{[Bor96]}
{I.N. Borzov, A.A. Fayans, E. Kr\"omer, and D. Zawischa, Z. Phys. {\bf A355},
  117 (1996)}.

\bibitem{[Dob94]}
{J. Dobaczewski, I. Hamamoto, W. Nazarewicz, and J.A. Sheikh, Phys. Rev. Lett.
  {\bf 72}, 981 (1994)}.

\bibitem{[Dec80]}
{J. Decharg\'e and D. Gogny, Phys. Rev. {\bf C21}, 1568 (1980)}.

\bibitem{[Nad94]}
{E.G. Nadjakov, K.P. Marinova, and Yu.P. Gangrsky, At. Data Nucl. Data Tables
  {\bf 56}, 133 (1994)}.

\bibitem{[Bat89]}
{C.J. Batty, E. Friedman, H.J. Gills, and H. Rebel, Adv. Nucl. Phys. {\bf 19},
  1 (1989)}.

\bibitem{[Rii92]}
{K. Riisager, A.S. Jensen, and P. M{\o}ller, Nucl. Phys. {\bf A548}, 393
  (1992)}.

\bibitem{[Lan64]}
{A.M. Lane, {\sl Nuclear Theory} (Benjamin, New York, 1964)}.

\bibitem{[Bru60]}
{K.A. Brueckner, T. Soda, P. W. Anderson, and P. Morel, Phys. Rev. {\bf 118},
  1442 (1960)}.

\bibitem{[Eme60]}
{V.J. Emery and A.M. Sessler, Phys. Rev. {\bf 119}, 248 (1960)}.

\bibitem{[Del95]}
{D.S. Delion, M. Baldo, and U. Lombardo, Nucl. Phys. {\bf A593}, 151 (1995)}.

\bibitem{[Kuc91]}
{H. Kucharek and P. Ping, Z. Phys. {\bf A339}, 23 (1991)}.

\bibitem{[Kad87]}
{S.G. Kadmenski\u{\i}, P.A. Luk'yanovich, Yu. I. Remesov, and V.I. Furman, Sov.
  J. Nucl. Phys. {\bf 45}, 585 (1987)}.

\bibitem{[Sap65]}
{\'E.E. Sapershte\u{\i}n and M.A. Troitski\u{\i}, Sov. J. Nucl. Phys. {\bf 1},
  284 (1965)}.

\bibitem{[Zaw87]}
{D. Zawischa, U. Regge, and R. Stapel, Phys. Lett. {\bf 185B}, 299 (1987)}.

\bibitem{[Reg88]}
{U. Regge and D. Zawischa, Phys. Rev. Lett. {\bf 61}, 149 (1988)}.

\bibitem{[Fay94]}
{S.A. Fayans, S.V. Tolokonnikov, E.L. Trykov, and D. Zawischa, Phys. Lett. {\bf
  338B}, 1 (1994)}.

\bibitem{[Bul80]}
{A. Bulgac, Preprint FT-194-1980, Central Institute of Physics, Bucharest,
  1980}.

\bibitem{[Bel87]}
{S.T. Belyaev, A.V. Smirnov, S.V. Tolokonnikov, and S.A. Fayans, Sov. J. Nucl.
  Phys. {\bf 45}, 783 (1987)}.

\bibitem{[Von91]}
{B.E. Vonderfecht, C.C. Gearhart, W.H. Dickhoff, A. Polls, and A. Ramos, Phys.
  Lett. {\bf B253}, 1 (1991)}.

\bibitem{[Deb94]}
{F. De Blasio and G. Lazzari, Nuovo Cimento {\bf 107A}, 1549 (1994)}.

\bibitem{[Ton78]}
{F. Tondeur, Z. Phys. {\bf A288}, 97 (1978)}.

\bibitem{[Hae89]}
{P. Haensel, J.L. Zdunik, and J. Dobaczewski, Astron. Astrophys. {\bf 222}, 353
  (1989)}.

\bibitem{[Smo93]}
{R. Smola\'nczuk and J. Dobaczewski, Phys. Rev. {\bf C48}, R2166 (1993)}.

\bibitem{[Che95]}
{B. Chen, J. Dobaczewski, K.-L. Kratz, K. Langanke, B. Pfeiffer, F.-K.
  Thielmann, and P. Vogel, Phys. Lett. {\bf B355}, 37 (1995)}.

\bibitem{[Pea96]}
{J.M. Pearson, R.C. Nayak, and S. Goriely, Phys. Lett. {\bf B387}, 455 (1996)}.

\bibitem{[Cha95a]}
{E. Chabanat, {\sl Interactions effectives pour des conditions extr\^emes
  d'isospin}, Universit\'e Claude Bernard Lyon-1, Thesis 1995, LYCEN~T~9501,
  unpublished}.

\bibitem{[Pat96]}
{Z. Patyk, A. Baran, J.F. Berger, J. Decharg\'e, J. Dobaczewski, R.
  Smola\'nczuk, and A. Sobiczewski, Acta Phys. Pol. {\bf B27}, 457 (1996)}.

\bibitem{[Cho95]}
{W.-T. Chou, R.F. Casten, and N.V. Zamfir, Phys. Rev. {\bf C51}, 2444 (1995)}.

\bibitem{[Det79]}
{C. D\'etraz, D. Guillemaud, G. Huber, R. Klapish, M. Langevin, F. Naulin, C.
  Thibault, L.C. Carraz, and F. Touchard, Phys. Rev. {\bf C19}, 164 (1979)}.

\bibitem{[Tou82]}
{F. Touchard, J. M. Serre, S. B\"uttenbach, P. Guimbal, R. Klapisch, M. de
  Saint Simon, C. Thibault, H. T. Duong, P. Juncar, S. Libermen, J. Pinard, and
  J. L. Vialle, Phys. Rev. {\bf C25}, 2756 (1982)}.

\bibitem{[Sch96]}
{H. Scheit, T. Glasmacher, B.A. Brown, J.A. Brown, P.D. Cottle, P.G. Hansen, R.
  Harkewicz, M. Hellstrom, R.W. Ibbotson, J.K. Jewell, K.W. Kemper, D.J.
  Morrissey, M. Steiner, P. Thirolf, and M. Thoennessen, Phys. Rev. Lett. {\bf
  77}, 3967 (1996)}.

\bibitem{[Gla97]}
{T. Glasmacher, B.A. Brown, M.J. Chromik, P.D. Cottle, M. Fauerbach, R.W.
  Ibbotson, K.W. Kemper, D.J. Morrissey, H. Scheit, D.W. Sklenicka, and M.
  Steiner, Preprint MSUCL-1048, 1977}.

\bibitem{[Cam75]}
{X. Campi, H. Flocard, A. K. Kerman, and S. Koonin, Nucl. Phys. {\bf A251}, 193
  (1975)}.

\bibitem{[Wer94a]}
{T.R. Werner, J.A. Sheikh, W. Nazarewicz, M.R. Strayer, A.S. Umar, and M. Misu,
  Phys. Lett. {\bf B333}, 303 (1994)}.

\bibitem{[Aud93]}
{G. Audi and A.H. Wapstra, Nucl. Phys. {\bf A565}, 1 (1993); Nucl. Phys. {\bf
  A565}, 66 (1993)}.

\bibitem{[Kra93]}
{K.-L. Kratz, J.-P. Bitouzet, F.-K. Thielemann, P. M\"oller, and B. Pfeiffer,
  Astrophys. J. {\bf 403}, 216 (1993)}.

\bibitem{[How93]}
{W.M. Howard, S. Goriely, M. Rayet, and M. Arnould, Astrophys. J. {\bf 417},
  713 (1993)}.

\bibitem{[Pfe97]}
{B. Pfeiffer, K.-L. Kratz, and F.-K. Thielemann, Z. Phys. {\bf A357}, 235
  (1997)}.

\bibitem{[Sur97]}
{R. Surman, J. Engel, J.R. Bennett, and B.S. Meyer, Preprint astro-ph/9701007}.

\bibitem{[Gor96]}
{S. Goriely and M. Arnould, Astron. Astrophys. {\bf 312}, 327 (1996)}.

\bibitem{[Hax97]}
{W.C. Haxton, K. Langanke, Y.Z. Qian, and P. Vogel, Phys. Rev. Lett. {\bf 78},
  2694 (1997)}.

\bibitem{[Qia97]}
{Y.Z. Qian, W.C. Haxton, K. Langanke, and P. Vogel, Phys. Rev. {\bf C55}, 1532
  (1997)}.

\bibitem{[Sta89]}
{A. Staudt, E. Bender, K. Muto, and H.V. Klapdor, Z. Phys. {\bf A334}, 47
  (1989)}.

\bibitem{[Mol90a]}
{P. M\"oller and J. Randrup, Nucl. Phys. {\bf A514}, 1 (1990)}.

\bibitem{[Tac95]}
{T. Tachibana and M. Arnould, Nucl. Phys. {\bf A588}, 333c (1995)}.

\bibitem{[Wig48]}
{E.P. Wigner, Phys. Rev. {\bf 73}, 1002 (1948)}.

\bibitem{[Uch85]}
{T. Uchiyama and H. Morinaga, Z. Phys. {\bf A320}, 273 (1985)}.

\bibitem{[Fay91]}
{S.A. Fayans, Phys. Lett. {\bf B267}, 443 (1991)}.

\bibitem{[Yok95]}
{M. Yokoyama, T. Otsuka, and N. Fukunishi, Phys. Rev. {\bf C52}, 1122 (1995)}.

\bibitem{[Sag95]}
{H. Sagawa, N. Van Giai, N. Takigawa, M. Ishihara, and K. Yazaki, Z. Phys. {\bf
  A351}, 385 (1995)}.

\bibitem{[Fuk91a]}
{M. Fukuda, T. Ichihara, N. Inabe, T. Kubo, H. Kumagai, T. Nakagawa, Y. Yano,
  I. Tanihata, M. Adachi, K. Asahi, M. Kouguchi, M. Ishihara, H. Sagawa, and S.
  Shimoura, Phys. Lett. {\bf B 268}, 339 (1991)}.

\bibitem{[Ram87]}
{S. Raman, C.H. Malarkey, W.T. Milner, C.W. Nestor, Jr., and P.H. Stelson,
  Atomic Data Nucl. Data Tables {\bf 36}, 1 (1987)}.

\bibitem{[Bou78]}
{M.Bouten, E. Flerackers, M.C. Bouten, Nucl. Phys. {\bf A307}, 413 (1978)}.

\bibitem{[Rag81]}
{I. Ragnarsson, S. \AA berg, H.-B. H\aa kansson and R.K. Sheline, Nucl. Phys.
  {\bf A361}, 1 (1981)}.

\bibitem{[Kan95]}
{Y. Kanada-Enyo, H. Horiuchi, A. Ono, Phys. Rev. {\bf C52}, 628 (1995)}.

\bibitem{[Esb95]}
{H. Esbensen, B.A. Brown, H. Sagawa, Phys. Rev. {\bf C51}, 1274 (1995)}.

\bibitem{[Vin95]}
{N. Vinh Mau, Nucl. Phys. {\bf A592}, 33 (1995)}.

\bibitem{[Ber95]}
{C.A. Bertulani, H. Sagawa, Nucl. Phys. {\bf A588}, 667 (1995)}.

\bibitem{[Li96]}
{X. Li and P.-H. Heenen, Phys. Rev. {\bf C54}, 1617 (1996)}.

\bibitem{[Mis97]}
{T. Misu, W. Nazarewicz, and S. {\AA}berg, Nucl. Phys. {\bf A614}, 44 (1997)}.

\bibitem{[Wer96]}
{T.R. Werner, J.A. Sheikh, M. Misu, W. Nazarewicz, J. Rikovska, K. Heeger, A.S.
  Umar, and M.R. Strayer, Nucl. Phys. {\bf A597}, 327 (1996)}.

\bibitem{[Nau96]}
{F. Naulin, P. Bonche, H. Flocard, and P.H. Heenen, Abstract at the {\it
  Conference on Nuclear Structure at the Limits}, Argonne 1996, ANL/PHY-96/1,
  p. 143}.

\bibitem{[Tho51]}
{R.G. Thomas, Phys. Rev. {\bf 81}, 148 (1951); {\bf 88}, 1109 (1952)}.

\bibitem{[Ehr51]}
{J.B. Ehrman, Phys. Rev. {\bf 81}, 412 (1951)}.

\bibitem{[Sch78a]}
{S. Schlomo, Rep. Prog. Phys. {\bf 41}, 957 (1978)}.

\bibitem{[Sch94]}
{R. Schneider, J. Friese, J. Reinhold, K. Zeitelhack, T. Faestermann, R.
  Gernhauser, H. Gilg, F. Heine, J. Homolka, P. Kienle, H.J. K\"orner, H.
  Geissel, G. M\"unzenberg, and K.Summerer, Z. Phys. A {\bf 348}, 241 (1994)}.

\bibitem{[Lew94]}
{M. Lewitowicz, R. Anne, G. Auger, D. Bazin, C. Borcea, V. Borrel, J.M. Corre,
  T. Dorfler, A. Fomichov, R. Grzywacz, D. Guillemaud-Mueller, R. Hue, M.
  Huyse, Z. Janas, H. Keller, S. Lukyanov, A.C. Mueller, Yu. Penionzhkevich, M.
  Pfutzner, F. Pougheon, K.Rykaczewski, M.G. Saint-Laurent, K. Schmidt, W.D.
  Schmidt-Ott, O. Sorlin, J. Szerypo, O. Tarasov, J. Wauters, and J. \.Zylicz,
  Phys. Lett. {\bf B332}, 20 (1994)}.

\bibitem{[Cha96]}
{M. Chartier, G. Auger, W. Mittig, A. Lepine-Szily, L.K. Fifield, J.M.
  Casandjian, M. Chabert, J. Ferme, A. Gillibert, M. Lewitowicz, M. Mac
  Cormick, M.H. Moscatello, O.H. Odland, N.A. Orr, G. Politi, C. Spitaels, and
  A.C.C. Villari, Phys. Rev. Lett. {\bf 77}, 2400 (1996)}.

\bibitem{[Gos64]}
{A. Goswami, Nucl. Phys. {\bf 60}, 228 (1964)}.

\bibitem{[Gos65]}
{A. Goswami and L.S. Kisslinger, Phys. Rev. {\bf 140 }, B26 (1965)}.

\bibitem{[Cam65]}
{P. Camiz, A. Covello, and M. Jean, Nouvo Cimento {\bf 36}, 663 (1965); {\bf
  42B}, 199 (1966)}.

\bibitem{[Bar69]}
{J. Bar-Touv, A. Gosvami, A.L. Goodman, and G.L. Struble, Phys. Rev. {\bf 178},
  178 (1969)}.

\bibitem{[Che78]}
{H.T. Chen, H. M\"uther, and A. Faessler, Nucl. Phys. {\bf A297}, 445 (1978)}.

\bibitem{[NP97]}
{W. Nazarewicz and S. Pittel, http://aps.org/BAPSAPR97/vpr/laye8.html}.

\bibitem{[Goo79]}
{A.L. Goodman, Adv. Nucl. Phys. {\bf 11}, 263 (1979)}.

\bibitem{[Zel96]}
{N. Zeldes, in Handbook of Nuclear Properties, Ed. by D. Poenaru and W.
  Greiner, Clarendon Press, Oxford, 1996, p. 13}.

\bibitem{[Che93]}
{J.M.C. Chen, J.W. Clark, R.D. Dav\'e, and V.V. Khodel, Nucl. Phys. {\bf A555},
  59 (1993)}.

\bibitem{[Che93a]}
{M.K. Cheoun, A. Bobyk, A. Faessler, F. Simkovic, and G. Teneva, Nucl. Phys.
  {\bf A561}, 74 (1993)}.

\bibitem{[Sch96a]}
{J. Schwieger, F. Simkovic, and A. Faessler, Nucl. Phys. {\bf A600}, 179
  (1996)}.

\bibitem{[Eng97]}
{J. Engel, S. Pittel, M. Stoitsov, P. Vogel, and J. Dukelsky, Phys. Rev. {\bf
  C55}, 1781 (1997)}.

\bibitem{[Fro70]}
{P. Fr\"obrich, Z. Phys. {\bf 236}, 153 (1970)}.

\bibitem{[Fro71]}
{P. Fr\"obrich, Phys. Lett. {\bf 37B}, 338 (1971)}.

\bibitem{[Kva90]}
{J. Kvasil, A.K. Jain, and R.K. Sheline, Czech. J. Phys. {\bf 40}, 278 (1990)}.

\bibitem{[Sat97a]}
{W. Satu{\l}a and R. Wyss, Phys. Lett. {\bf B393}, 1 (1997)}.

\bibitem{[Pal75]}
{F. Palumbo, Lett. Nuovo Cimento {\bf 14}, 572 (1975)}.

\bibitem{[Alm90]}
{T. Alm, G. R\"opke, and M. Schmidt, Z. Phys. {\bf 337}, 355 (1990)}.

\bibitem{[Abo95]}
{Y. Aboussir, J.M. Pearson, A.K. Dutta, and F. Tondeur, At. Data Nucl. Data
  Tables {\bf 61}, 127 (1995)}.

\bibitem{[Sat97]}
{W. Satu{\l}a, D.J. Dean, J. Gary, S. Mizutori, and W. Nazarewicz, Phys. Lett.
  {\bf B}, in press}.

\bibitem{[Bre90]}
{D.S. Brenner, C. Wesselborg, R.F. Casten, D.D. Warner and J.-Y. Zhang, Phys.
  Lett. {\bf 243B}, 1 (1990)}.

\bibitem{[Eng96]}
{J. Engel, K. Langanke, and P. Vogel, Phys. Lett. {\bf B389}, 211 (1996)}.

\bibitem{[Per96]}
{E. Perli\'nska, S.G. Rohozi\'nski, J. Dobaczewski, and W. Nazarewicz, Proc. of
  Int. Hirschegg Workshop XXIV, {\sl Extremes of Nuclear Structure} edited by
  H. Feldmeier, J. Knoll, and W. N\"orenberg, (GSI, Darmstadt, 1996), p. 228}.

\bibitem{[Goo72]}
{A.L. Goodman, Nucl. Phys. {\bf A186}, 475 (1972)}.

\bibitem{[Wol71]}
{H. Wolter, A. Faessler, and P. Sauer, Phys. Lett. {\bf 31B}, 516 (1970); Nucl.
  Phys. {\bf A167}, 108 (1971)}.

\bibitem{[Hof95b]}
{S. Hofmann, Radiochim. Acta {\bf 70/71}, 93 (1995)}.

\bibitem{[Hof97]}
{S. Hofmann, in: {\em Nuclear Decay Modes}, ed. by D.N. Poenaru and W. Greiner,
  (IOP, Bristol, 1996); Preprint GSI-93-04, 1993}.

\bibitem{[Kad71]}
{S.G. Kadmenski\u{\i} and V.E. Kalechtis, Sov. J. Nucl. Phys. {\bf 12}, 37
  (1971)}.

\bibitem{[Fei83]}
{W.F. Feix and E.R. Hilf, Phys. Lett. {\bf 120B}, 14 (1983)}.

\bibitem{[Abe97]}
{S. \AA berg, P.B. Semmes, and W. Nazarewicz, Phys. Rev. {\bf C}, October
  1997}.

\bibitem{[Gur88]}
{S.A. Gurvitz, Phys. Rev. {\bf A38}, 1747 (1988)}.

\bibitem{[Buc92]}
{B. Buck, A.C. Merchant, and S.M. Perez, Phys. Rev. {\bf C45}, 1688 (1992)}.

\bibitem{[Kad96]}
{S.G. Kadmenski\u{\i} and V.P. Bugrov, Phys. Atomic Nuclei {\bf 59}, 399
  (1996)}.

\bibitem{[ISL91]}
{{\em The IsoSpin Laboratory: Research opportunities with Radioactive Nuclear
  Beams}, Preprint LALP 91-51}.

\end{thebibliography}

\end{document}